\documentclass[aps,prx,reprint,twocolumn,amsmath,amssymb,superscriptaddress]{revtex4-2}
\usepackage{braket}

\usepackage{amsmath,amsfonts,amssymb,bm,bbm}
\usepackage{graphicx}
\usepackage{hyperref}
\hypersetup{
    colorlinks=true,
    citecolor=magenta,
    linkcolor=blue,    
    urlcolor=blue,
}
\usepackage{physics}
\usepackage{dsfont}
\usepackage{bbold}
\usepackage{soul}
\usepackage{enumitem}
\usepackage[table]{xcolor}
\usepackage{helvet}

\definecolor{purple}{HTML}{9326ff}

\newcommand{\figpanel}[1]{\textbf{\textsf{#1}}}

\definecolor{light_blue}{HTML}{f0f5ff}
\definecolor{light_grey}{HTML}{ededed}
\definecolor{check}{HTML}{ff7300}
\definecolor{energy_orange}{HTML}{ffb061}

\usepackage{titlesec}
\titleformat*{\subsection}
    {\fontsize{10}{10}\bfseries}
    {}

\titlespacing*{\subsection}
  {0pt}         
  {15pt}        
  {5pt}       
  
\titleformat*{\subsubsection}%
    {\fontsize{10}{10}\itshape}%
    {}

\titlespacing*{\subsubsection}
  {0pt}         % <left> (no extra left margin)
  {15pt}        % <before-sep> (space above the heading)
  {5pt}         % <after-sep> (space below the heading)





\newcommand{\methods}{\hyperlink{methods}{Methods}}

\begin{document}

\title{Superextensive learning in quantum reservoirs \texorpdfstring{\\}{ } at the onset of information scrambling}

\author{Jonas Freiheit}
\affiliation{Department of Physics, School of Science, RMIT University, Melbourne, 3000, Victoria, Australia}
\affiliation{RMIT Applied Quantum Technologies Centre, RMIT University, Melbourne, 3000, Victoria, Australia}

\author{Francesco Campaioli}
\email{francesco.campaioli@rmit.edu.au}
\affiliation{Department of Physics, School of Science, RMIT University, Melbourne, 3000, Victoria, Australia}
\affiliation{RMIT Applied Quantum Technologies Centre, RMIT University, Melbourne, 3000, Victoria, Australia}

\date{\today}

\begin{abstract}
The idea that information processing is optimised near the boundary between order and chaos has emerged as a recurring principle across neuroscience, complex systems, and machine learning. 
Here we test this hypothesis in quantum many-body systems, numerically simulating two-dimensional Ising networks of up to $N=20$ spins, operated as quantum reservoirs for time-series forecasting. 
Using out-of-time-order correlators (OTOCs), we locate the onset of information scrambling as the input strength is swept, separating regimes where information is \textit{frozen} and \textit{scrambled} across the whole reservoir state. 
We show that prediction precision peaks at the onset of scrambling, along with the number of computational-basis states that the reservoir actively populates. We then show that prediction precision grows as a power law $\sim N^{\alpha}$ in the reservoir size, superextensively $(\alpha>1)$ at the onset of scrambling and only sublinearly $(\alpha<1)$ in either neighbouring regime.
Finally, we show that scrambling enhances the nonlinear components of the reservoir memory while reducing its linear capacity. At the onset, the total memory capacity grows superextensively, provided that the necessary ``forgetting'' mechanism is supplied by a collective relaxation channel.
These results consolidate the role of information scrambling in learning systems, turning it from an operating point into a scaling law for the performance of quantum reservoirs.
\end{abstract}

\maketitle

\noindent
Studies in neuroscience, neuromorphic computing and machine learning have converged on a common principle: a system processes information best close to a critical point separating two phases in the operating regime of its substrate~\cite{Beggs2003,Shew2009,Munoz2018,Stieg2012,Hochstetter2021,Tanaka2019,langton1990,Bertschinger2004,Lukosevicius2009}. In neuroscience, the critical-brain hypothesis holds that cortical circuits self-organise to such a point, where dynamic range, information transmission, and memory are jointly maximised~\cite{Chialvo2010,Munoz2018}. Two distinct notions of criticality are invoked in its support. \emph{Avalanche} criticality places a system at the critical point of an activity-propagation transition, where a perturbation triggers cascades whose sizes and lifetimes are scale-free. It benefits systems held at a fixed operating point~\cite{Shew2009,Shew2011,ShewPlenz2013}, and its signature is the power-law statistics of avalanches~\cite{Beggs2003,Petermann2009,Shriki2013}. \emph{Edge-of-chaos} criticality instead concerns how systems driven by time-dependent inputs respond to perturbations, and identifies the regime in which perturbations are sustained rather than damped away or amplified out of control.~\cite{langton1990,Bertschinger2004,Legenstein2007}.

In neuromorphic computing, both notions appear in physical systems whose own dynamics carry out the computation. Self-organising nanowire and nanoparticle networks display avalanche statistics and a tunable order-to-chaos response~\cite{Stieg2012,Mallinson2019,Zhu2021}, with the latter being optimal for time-dependent learning tasks~\cite{Tanaka2019,Mallinson2019,Hochstetter2021}. Machine learning has given this principle a quantitative form through \textit{reservoir computing}: a time-dependent input stream is fed into a fixed, untrained dynamical system---the \emph{reservoir}---and its state is read out through a single trained layer, so that task performance is a property of the reservoir dynamics~\cite{Jaeger2004,Maass2002}. Performance peaks between the ordered and chaotic regimes, where the dynamics offer many active and mutually independent features~\cite{Dambre2012,Boedecker2012,GrigoryevaOrtega2018}.
\begin{figure*}[t]
  \centering
  \includegraphics[width=0.98\textwidth]{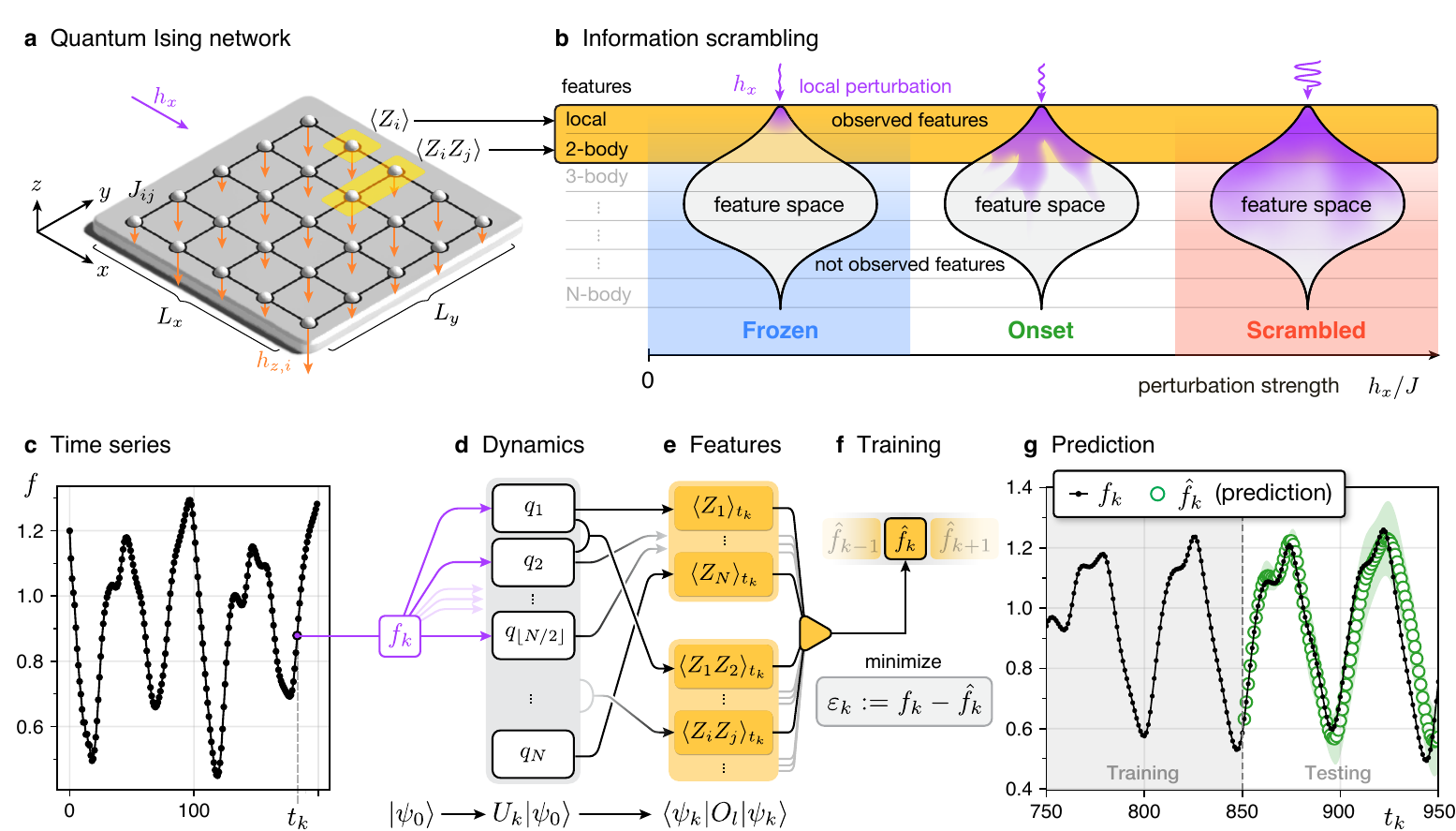}
    \caption{\textbf{Time-series prediction with quantum Ising networks, and the role of information scrambling.}
  \figpanel{a} We consider a reservoir of $N = L_x \times L_y$ spins on a two-dimensional lattice, coupled to their nearest neighbours by a weakly disordered Ising interaction $J_{ij}$, driven by a global transverse field $h_x$ and a weakly disordered local longitudinal field $h_{z,i}$. Only local $\langle Z_i\rangle$ and two-body $\langle Z_i Z_j\rangle$ correlations in the computational basis are measured to limit measurement and memory costs.
  \figpanel{b} A local perturbation encoded in the transverse field is applied to the system, spreading via the reservoir dynamics. When $h_x/J$ is small the reservoir is \textit{frozen} and the perturbation remains localised. When it is large the perturbation is scrambled almost entirely into many-body features that become self-similar. At the onset of scrambling, the perturbation populates a rich portion of observed features. The structure sketched here is quantified in Fig.~\ref{fig:activity}.
  \figpanel{c}--\figpanel{f} The protocol for time-series prediction. Each value $f_k = f(t_k)$ of a scalar input series $f(t)$ \figpanel{c} is written into half of the spins, here labelled as $q_1, \dots, q_{\lfloor N/2\rfloor}$, via a transverse field of proportional strength. The network evolves unitarily for a fixed interval \figpanel{d}, and the resulting state is measured in the computational basis to yield the features $\langle \psi_k | O_l | \psi_k \rangle$ \figpanel{e}. The network is then reset, so that memory of earlier inputs is supplied classically rather than held in the quantum state. Only a linear readout is trained, by ridge regression on these features \figpanel{f}, to minimise the error $\varepsilon_k$ between inputs $f_k$ and outputs $\hat{f}_k = \sum_l w_l\; m_{l,k}$ by varying the weights $w_l$ associated to the features $m_{l,k}$, and the reservoir itself is never optimised.
  \figpanel{g} The trained readout is fitted on a training interval and evaluated on a testing set, with prediction precision defined as the inverse of the normalised error over the test set.}
  \label{fig:overview}
\end{figure*}

Here, we explore how this principle translates to quantum systems through the lens of \textit{information scrambling}. Scrambling describes how information injected locally becomes encoded in correlations shared across an entire system, becoming irretrievable by any local measurement~\cite{Hayden2007,Sekino2008,Hosur2016}. It is the quantum-information counterpart of the sensitivity to initial conditions that defines classical chaos~\cite{MSS2016,Swingle2018,XuSwingle2024}, hypothesised to be underpinning information release from evaporating black holes~\cite{Hayden2007,MSS2016} and computational complexity in quantum circuits~\cite{RobertsYoshida2017,Dowling2023,Swingle2018,GoogleEchos}.

Scrambling is also, in essence, the mechanism underpinning reservoir computing. In quantum reservoir computing (QRC), an input signal is injected into a many-body quantum system and spread by its own dynamics, generating a rich feature map for learning tasks~\cite{Nakajima2019}. This framework, introduced to investigate the role of superposition and entanglement in machine learning~\cite{Fujii2017}, has become the object of intense fundamental and applied research~\cite{Mujal2021,Kobayashi2024, Sannia2024,Hu2024,Xiong2025, Cenedese2026}. Spin networks have drawn particular attention for their feasibility, across nuclear magnetic resonance~\cite{Negoro2018,Hou2026}, superconducting processors~\cite{Yasuda2023,Hu2024}, and Rydberg-atom arrays~\cite{Bravo2022,Kornjaca2024}. Recent work has shown that performance rises towards the thermalisation transition of a closed disordered spin network, which enables the fading memory required by learning~\cite{martinezpena2021}, and that quantum reservoirs can detect quantum phase transitions~\cite{Kobayashi2025probe}, distinguishing chaotic from non-chaotic dynamics~\cite{Kobayashi2026edge}. However, two questions remain open: can learning performance outpace operating costs as the reservoir grows, and is edge-of-chaos criticality needed to achieve that, or is information scrambling enough?

In this work, we address these questions by numerically simulating two-dimensional transverse-field Ising networks of up to $N=20$ spins, operated as quantum reservoirs for time-series prediction, as illustrated in Fig.~\ref{fig:overview}. A single dimensionless control parameter, measuring the strength of the transverse field relative to the spin--spin interactions, drives the dynamics from a \textit{frozen} to a \textit{scrambled} regime. To characterise this transition we use out-of-time-order correlators (OTOCs), the standard probe of information scrambling and quantum many-body chaos~\cite{MSS2016,Swingle2018,Dowling2023}, recently demonstrated on Google's superconducting processor~\cite{GoogleEchos}, as a path towards verifiable practical quantum advantage~\cite{bermejo2026tensor}.

First, we show that prediction precision peaks at the onset of information scrambling, where the number of computational-basis states expressed by the reservoir grows exponentially with $N$. There, precision grows superextensively with $N$ even when readout is restricted to local measurements and two-body correlations, far outpacing both frozen and scrambled regimes. 
Scrambling also lifts the expressive capacity, driving it towards the ceiling set by the number of measured features
and past the bound that applies to local input encoding in gate-based QRC~\cite{Hu2023REC,Schuld2021,Innocenti2023,Schutte2025}.
Finally, allowing the reservoir to retain its own quantum state between inputs, we show that scrambling promotes nonlinear memory capacity, hindering linear memory. At the onset of scrambling, the total memory capacity grows superextensively, as long as the required fading memory is provided by a collective relaxation channel rather than a local one. 

Taken together, these results establish the onset of scrambling as an operating principle for quantum reservoirs, indicating that it can survive at scale under tractable operating costs. 

% --- results ---
\subsection*{Results}
 
\subsubsection*{Time-series prediction with quantum reservoirs}

\noindent
Let us begin by introducing the quantum reservoir computing framework used throughout, leaving details to \methods{}. We consider reservoirs given by a network of $N = L_x \times L_y$ spins on a two-dimensional rectangular lattice, defined by the time-independent transverse field Ising Hamiltonian $H_0$
\begin{equation}
H_0 = \sum_{\langle i,j\rangle} J_{ij}\, Z_i Z_j
  + h_x \sum_i X_i
  + \sum_i h_{z,i}\, Z_i ,
\label{eq:H}
\end{equation}
and illustrated in Fig.~\ref{fig:overview}~\figpanel{a}, where the sum runs over nearest-neighbours pairs $\langle i,j\rangle$ in the square lattice, where $X_i$ and $Z_i$ represent the $x$ and $z$ Pauli operator on spin $i$, respectively. Nearest neighbour pairs $i,j$ are coupled by an Ising interaction $J_{ij}$ of mean strength $J>0$, and driven by a global transverse field $h_{x}$ and a local longitudinal field $h_{z,i}$ that acts along the coupling axis. Weak disorder in the couplings and in the longitudinal fields renders the dynamics generically non-integrable, leaving the ratio $h_x/J$ as the single control parameter, in a regime accessible to neutral-atom and superconducting platforms~\cite{Bravo2022,Yasuda2023,Kornjaca2024}; see \methods{} for details.

We focus on time-series prediction, and consider the history of a scalar signal $f(t)$ sampled at evenly spaced time-steps $t_k$, with $\Delta t = t_{k+1}-t_{k}$, with the aim of forecasting its value some number of steps ahead. Each input $f_k = f(t_k)$ is turned by the network into a set of real-valued features $m_{l,k}$, from which a trained linear readout reconstructs the target as a weighted sum, $\hat{f}_k = \sum_l w_l\, m_{l,k} \approx f_k$. Only the weights $w_l$ are fitted, separately for each forecast horizon; the network itself---i.e., its Hamiltonian---is never optimised.

At each step $k$, the input enters as an additional transverse-field driving term acting on half of the spins, modulated as $h_x \to h_x(1+f_k)$ on that subset. The network evolves from a fiducial state $\ket{\psi_0}$ for a fixed interval and is measured; see Eqs.~\eqref{eq:Hk}--\eqref{eq:unitary} and \methods{} for details. The encoding is therefore linear in the input and local in the network, and because it addresses a fixed fraction of the spins with a single field amplitude, its cost does not grow with $N$.
\begin{figure}
  \centering
  \includegraphics[width=0.48\textwidth]{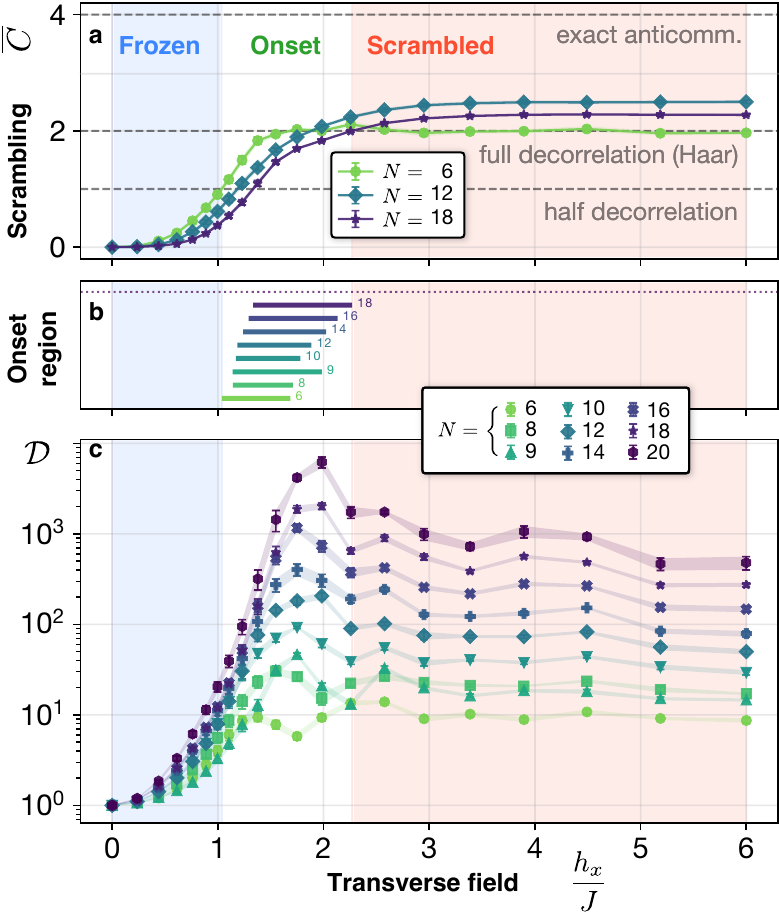}
\caption{\textbf{Information scrambling and computational basis states}
  \figpanel{a} OTOC saturation $\overline{C}$, defined in Eq.~\eqref{eq:Cbar}, versus the transverse field $h_{x}/J$, for three network sizes $N = 6, 12, 18$. $\overline{C}$ measures how completely a local perturbation has decorrelated from its initial support, with dashed lines marking half decorrelation ($\overline{C}=1$), full decorrelation ($\overline{C}=2$), and exact anticommutation ($\overline{C}=4$).
  \figpanel{b} The scrambling onset region for each size $N$, bounded by the fields at which $\overline{C}$ crosses $1$ and $2$. The window width and position depend on network size and geometry. The frozen ($\overline{C}<1$, blue) and scrambled ($\overline{C}>2$, red) regions are shown with coloured shading in every panel, marking the fields lying below and above every window.
  \figpanel{c} Participation ratio $\mathcal{D}$, of Eq.~\eqref{eq:pratio}, measuring the number of computational-basis states carrying appreciable weight, on a logarithmic scale. At every size $\mathcal{D}$ peaks close to the upper edge of the scrambling window; its maximum grows exponentially in $N$, by more than two orders of magnitude between $N=6$ and $N=20$. Error bars and shaded areas denote standard deviation over disorder realisations.}
  \label{fig:OTOC-saturation}
\end{figure}

We limit the readout to single-spin and two-spin observables in the computational basis, so that the raw features
\begin{equation}
r_{l,k} = \langle \psi_k | O_l | \psi_k \rangle ,
\qquad O_l \in \{ Z_i \} \cup \{ Z_i Z_j \} ,
\label{eq:features}
\end{equation}
comprise the magnetisation of each spin and the correlation of each pair, with the indices $i$ and $j$ running over the whole network. Note that we deliberately reset the reservoir state at each step to align with the most feasible operating approach available on current platforms~\cite{Ebadi2021,Scholl2021,Krantz2019,Johnson2011}. This approach requires neither coherent memory between inputs nor mid-circuit feedback, offering an operating mode executable on available devices. Since reset erases memory encoded in the reservoir states, we restore temporal context classically, by making the final features $m_{l,k}$ depend on the history of the raw features $r_{l,k}$, following the quantum--classical hybrid approach of Ref.~\cite{Settino2025}; see~\methods{}. 
In the final section we lift this restriction, retaining the quantum state between steps to examine the role of decoherence and scrambling on memory.

Two costs are worth separating. The cost of the exact numerical simulations used in this work is dominated by feature generation, with memory and operations scaling exponentially with $N$. However, in practice, i.e., on real devices, the cost depends on encoding and measurement. For the framework considered here, estimating computational-basis observables to an error $\eta$ requires a number of shots that grows with $\eta^{-2}$ and logarithmically in $N$; see~\methods{}. For post-processing and learning, handling and storing features scales with their number; we keep this cost polynomial in $N$ by design. Here we compute and analyse noiseless features obtained in the infinite-shot limit; shot-limited estimates can be derived following Ref.~\cite{Schutte2025}. 

% --- scrambling location ----
\subsubsection*{Locating the edge of information scrambling}
\noindent
We now look at the networks' ability to scramble information, measured by how rapidly a perturbation applied to a spin affects distant ones. This is quantified by out-of-time-order correlators~\cite{MSS2016,Swingle2018}: for two spins $i$ and $j$ we evaluate
\begin{equation}
C_{ij}(t) = \big\lVert \big[Z_i(t), Z_j \big] \ket{\psi_0} \big\rVert^{2},
\label{eq:otoc}
\end{equation}
with $Z_i(t) = U_0^{\dagger}(t)\, Z_i\, U_0^{\phantom{\dagger}}(t)$ and $U_0(t) = \exp(-i H_{0} t)$. The correlators $C_{ij}(t)$ start from zero and grow as the perturbation travels from $i$ to $j$, saturating at a value that reflects how thoroughly the information has been scattered across the network. Let us note that, while this work was being completed, Ref.~\cite{Keenan2026} independently identified scrambling as a governing quantity in quantum reservoirs, through an information-theoretic analysis of storage and information loss~\cite{Keenan2026}, presenting results for $N=6$. Here, we take a different route and look at the saturation of out-of-time-order correlators, which provides an absolute reference against which the operating point can be calibrated, and we use it to ask how the optimum behaves as the network grows.

We then express information scrambling with a single number by evaluating the OTOC saturation $\overline{C} \in [0,4]$, given by the late-time value of $C_{ij}$ averaged over all spin pairs with $i=1$ selected as the input site; see Eq.~\eqref{eq:Cbar} in \methods{}. Its value is indicative of distinct regimes: $\overline{C} = 0$ when the perturbation and the measurement still commute, so that information remains frozen where it was injected, and $\overline{C} = 2$ when they have decorrelated completely, which is also the value a random evolution would produce~\cite{Hosur2016,RobertsYoshida2017,Fan2017}, while a value of $\bar{C}=4$ corresponds to exact anticommutation.

Using $\overline{C}$ we divide the control parameter $h_x/J$ sweep into regimes without needing reference to any task, as shown in Fig.~\ref{fig:OTOC-saturation}~\figpanel{a} and~\figpanel{b}. The network dynamics is \textit{frozen} where $\overline{C} < 1$ and \textit{scrambled} where $\overline{C} > 2$. The scrambling onset lies between these two regions. It is important to notice that, although the mixed-field Ising model is established as non-integrable and thermalising at comparable parameters~\cite{Banuls2011,KimHuse2013}, saturation of $\overline{C}$ certifies scrambling, which is a necessary but not a sufficient condition for chaos~\cite{Dowling2023}. We therefore use $\overline{C}$ to locate the boundary between the frozen and scrambled regimes, rather than as an order parameter.
 
To further interrogate the role of the transverse field, we ask how widely the reservoir state spreads over the basis in which we actually measure it. We do so by evaluating the \textit{participation ratio} of the reservoir state
\begin{equation}
\mathcal{D} = \Big( \sum_b |\braket{b}{\psi}|^{4} \Big)^{-1} ,
\label{eq:pratio}
\end{equation}
where the sum runs over the computational basis $\{\ket{b}\}$~\cite{Tsukerman2017}. The participation ratio counts how many basis states carry appreciable weight, with $\mathcal{D} = 1$ when the network occupies a single basis state, growing as the distribution spreads. Thus, a large value of $\mathcal{D}$ indicates that the reservoir states can supply a large number of distinguishable accessible features.
 
Sweeping the transverse field, $\mathcal{D}$ rises steeply through the scrambling onset window, peaking sharply at its upper edge, and then falling significantly once the network is fully scrambled, as shown in Fig.~\ref{fig:OTOC-saturation}~\figpanel{c} for $N=6$---$20$. The state is spread most widely over the measurable basis at the onset of scrambling. Deeper in the scrambled regime the dynamics are increasingly dominated by the local transverse field, becoming weakly correlated.

% --- expressivity ----
\subsubsection*{Scrambling lifts expressive capacity}
\noindent
Before looking at predictive performance, we study how scrambling affects the ability of the reservoirs to express independent functions of the input. This is quantified by the \textit{expressive capacity} (EC), the number of linearly independent functions of the input that a system can supply through its readout~\cite{Dambre2012, Hu2023REC}. Here, we evaluate the EC in the noiseless, infinite-shot limit, where it reduces to the rank of the Gram matrix of the features; see Eq.~\eqref{eq:expressive_capacity} in~\methods{}. At a finite shot budget, only the directions rising above the sampling noise remain distinguishable, and the resolvable expressive capacity (REC) follows from the same spectrum once a signal-to-noise threshold is imposed~\cite{Hu2023REC,Schutte2025}; see \methods{} for details. 
\begin{figure*}[t]
  \centering
  \includegraphics[width=0.98\textwidth]{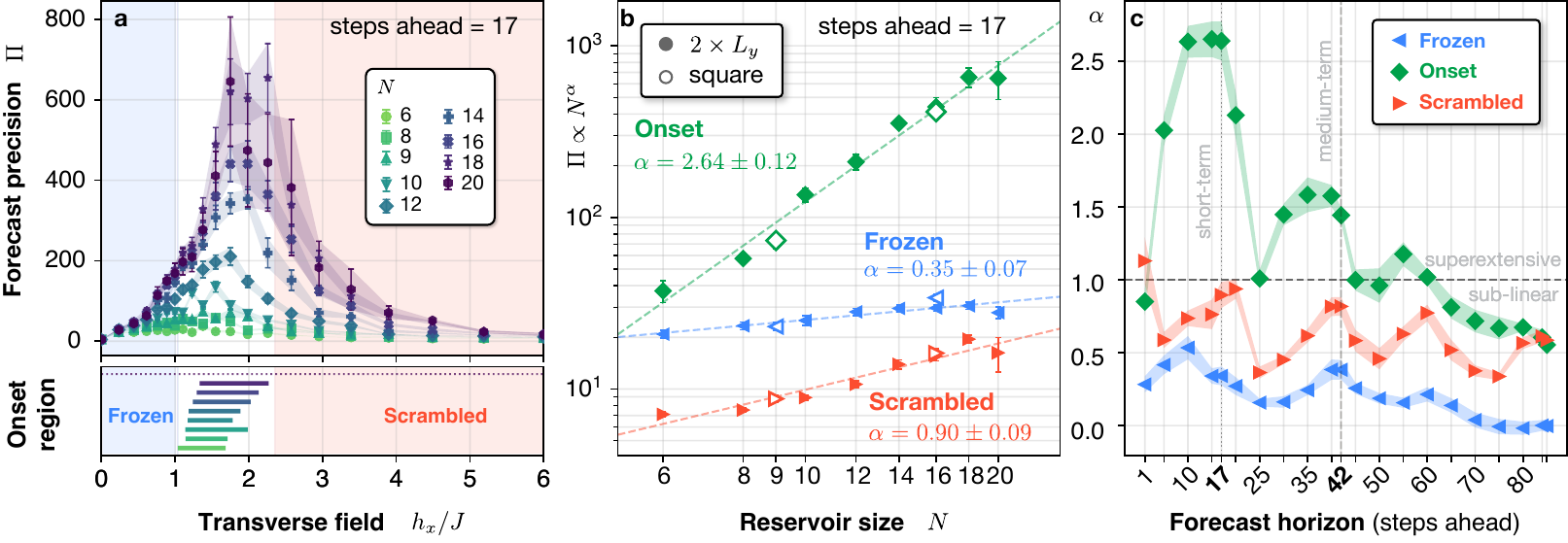}
\caption{\textbf{Superextensive predictive precision at the onset of scrambling.}
  \figpanel{a} Forecasting precision $\Pi = \varepsilon^{-1}$, with $\varepsilon$ the normalised root-mean-square error between target and prediction of Eq.~\eqref{eq:precision}, versus the transverse field $h_{x}/J$ at a forecast horizon of $17$ steps ahead, matching the intrinsic timescale of the Mackey--Glass task at $\tau_\mathrm{MG}=17$, for network sizes $N=6$ to $20$. The lower strip reproduces the scrambling onset region for each $N$.
  \figpanel{b} Precision at the same horizon as a function of reservoir size $N$ on logarithmic axes, evaluated at its peak within the onset region and at representative frozen and scrambled fields, $h_x/J = 0.237$ and $h_x/J = 6.00$, with power-law fits $\Pi \propto N^{\alpha}$. Precision grows superextensively at the peak, $\alpha = 2.64 \pm 0.12$, and sublinearly in both neighbouring regions: $\alpha = 0.35 \pm 0.07$ in the frozen regime and $\alpha = 0.90 \pm 0.09$ in the scrambled regime. Rectangular ($2\times L_y$, filled markers) and square (open markers) lattices obey the same scaling, indicating that the exponent is set by the number of spins rather than by the geometry of the network.
  \figpanel{c} The fitted exponent in each region as a function of forecast horizon, resolved over the full range from $1$ to $85$ steps ahead, with $17$ and $42$ highlighted as standard short- and mid-term benchmarks. The dashed line at $\alpha = 1$ separates superextensive from sublinear scaling; the vertical dotted line marks the short-term horizon of \figpanel{a} and \figpanel{b}, while the vertical dashed line marks the medium-term horizon of 42 steps ahead. The advantage at the onset is largest at short horizons, persists at intermediate ones, and is lost at the longest, where all three regions scale sublinearly. Shaded areas denote the uncertainty on $\alpha$.}
  \label{fig:precision-scaling}
\end{figure*}

Evaluating the expressive capacity across the sweep, we find that it tracks the spreading of information, following a power law $\propto N^\alpha$ of the network size $N$; see Eq.~\eqref{eq:shot-based-rec} in ~\methods{} and Fig.~\ref{fig:rec} for details. Expressive capacity grows sublinearly ($\alpha=0.75$ at $h_x/J = 0.25$) in the frozen region, and superextensively both at the onset ($\alpha=1.42$ averaged in the onset window $h_x/J\in[1,2]$) and in the scrambled region ($\alpha=1.72$ at $h_x/J = 6.00$). 
The resulting dynamics spread the input across the network, lifting expressivity above the linear Hamming bound, which limits common-weight single-qubit encoding in gate-based implementations~\cite{Schutte2025}, and towards the maximum number $\mathcal{R}_2 = N(N+1)/2$ of independent features available from local and two-body observables in the computational basis, given in Eq.~\eqref{eq:featurecount}.
 
Note that expressive capacity is a necessary but not sufficient condition for predictive performance, since it offers only an upper bound on what a reservoir can compute. Indeed, the largest exponent obtained here belongs to the scrambled region, where time-series prediction performance is poor, as we discuss in the next section by evaluating precision explicitly.

% --- precision ----
\subsubsection*{Superextensive predictive precision}

\noindent
Let us now look at the forecasting precision of the networks. We benchmark predictive performance on the Mackey--Glass (MG) series~\cite{mackey1977oscillation}, a standard test, whose chaotic dynamics are governed by a delay $\tau_{\mathrm{MG}}$ that sets the intrinsic timescale of the signal; see Eq.~\eqref{eq:mackeyglass} in \methods{}. Here, we take $\tau_{\mathrm{MG}} = 17$, the canonical choice for time-series prediction benchmarks, which places the series well inside its chaotic regime. All forecasts are \textit{open-loop} and direct, meaning that the network is always driven by the target MG time series, training separate readout weights for different forecast horizons. Outputs $\hat{f}_k$ are never fed back into the reservoir, so errors are not compounded over longer horizons. Performance is quantified by the precision $\Pi = \varepsilon^{-1}$, the reciprocal of the normalised root-mean-square error (NRMSE) $\varepsilon$ between the target $f_k$ and the prediction $\hat{f}_k$ over a test set, after training of the readout layers; see Eq.~\eqref{eq:precision} for details. Note that $\Pi$ is dimensionless and larger values indicate better forecasts.
 
Sweeping the control parameter, precision peaks at the onset of scrambling, close to the upper edge of the region, as shown in Fig.~\ref{fig:precision-scaling}~\figpanel{a} for all considered size from $N = 6$ to $N=20$ at a forecast horizon of $17$ steps, matching the intrinsic timescale of the series. Importantly, the height of the precision peak increases with $N$, as shown in Fig.~\ref{fig:precision-scaling}~\figpanel{a} and~\figpanel{b}. At the upper edge of the onset region, precision grows as a power law $\Pi \propto N^{\alpha}$ with $\alpha = 2.64 \pm 0.12$. The same fit gives $\alpha = 0.35 \pm 0.07$ in the frozen region and $\alpha = 0.90 \pm 0.09$ in the scrambled one. Rectangular and square lattices fall on the same line, so the exponent reflects the number of spins rather than the aspect ratio of the network. These results show that the performance at the onset of scrambling scales superextensively over the range of sizes accessed here, and that the separation from the neighbouring regimes grows as spins are added. 
\begin{figure*}[t]
    \centering
    \includegraphics[width=0.98\textwidth]{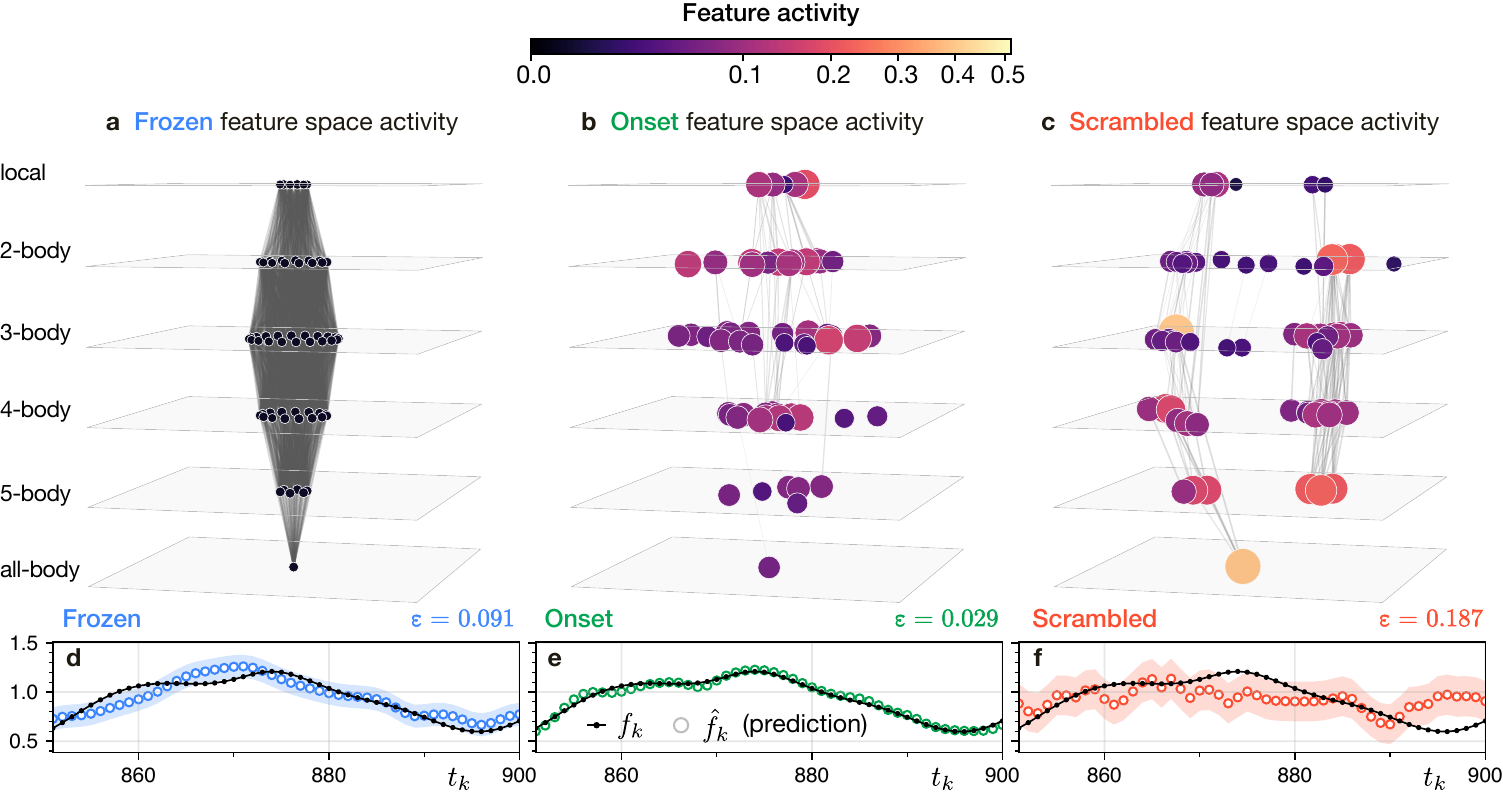}
    \caption{\textbf{Feature-space activity in the three regimes.}
    \figpanel{a}--\figpanel{c} The feature space of a $2\times3$ reservoir ($N=6$), at $h_x/J = 0.05$ (frozen), $1.75$ (onset) and $6.00$ (scrambled), represented as a graph. All $63$ computational-basis correlators are shown, stratified into layers of $s$-body operators; node colour and size give the activity of a feature over the run, per Eq.~\eqref{eq:activity}, and an edge joins any pair carrying almost identical information, as per Eq.~\eqref{eq:featcorr}. In-plane node positions are chosen so that similar features are drawn together both within and across layers. A layer whose features are mutually degenerate carries no relative geometry and is drawn as an evenly spaced ring.
    \figpanel{d}--\figpanel{f} Forecasts of the Mackey--Glass series over a common test window at a horizon of $15$ steps, trained as everywhere else on the local and two-body features alone (top two layers), with classical memory optimised at that horizon. The prediction error $\varepsilon$, quantified by the NRMSE of Eq.~\eqref{eq:precision}, is lowest at the onset, where feature degeneracy is broken and duplicates are fewest ($84$ pairs), followed by the scrambled region ($138$ pairs) and the frozen one ($1953$ pairs).}
    \label{fig:activity}
\end{figure*}

The scaling exponent depends on how far ahead the network is asked to forecast, as shown in Fig.~\ref{fig:precision-scaling}~\figpanel{c}. It is largest on short-term horizons of $10$--$20$ steps, comparable to the intrinsic timescale $\tau_{\mathrm{MG}} = 17$ of the series, where the exponent at the onset exceeds $2.5$. Scalability survives the mid-term horizon of $42$ steps, where precision still grows superextensively at the onset of scrambling, but fails at the long-term horizon of $84$ steps, where precision scales sublinearly in all three operative regions. Note that these two longer horizons are demanding benchmarks, for which performance also depends on the length of the training set; see~\methods{}. At the shortest horizon, by contrast, the task is close to linear and requires no scrambling, leading to similar scaling in all the three regions. Importantly, superextensive precision is available across the range of horizons over which forecasting a chaotic signal is both difficult and useful.

% --- activity ----
\subsubsection*{Activity in the feature space}
\noindent
Looking at the activity of features offers insights into the fall of precision scaling in the frozen and scrambled regions. Both failures are visible in Fig.~\ref{fig:activity}, which represents the feature space as a layered graph. Each node is a computational-basis correlator $r_{l,k}$, given in Eq.~\eqref{eq:features}, coloured and sized by its \textit{activity}, 
\begin{equation}
a_l = \sqrt{ \big\langle r_{l,k}^2 \big\rangle_k - \big\langle r_{l,k} \big\rangle_k^2 },
\label{eq:activity}
\end{equation} 
where $\langle{\cdots}\rangle_k$ represents the average over the time domain,
i.e., the standard deviation of its expectation value across the input series. Each pair of near-duplicate features is joined by an edge, which measures their interdependence, 
\begin{equation}
P_{ll'} = \frac{ \big\langle r_{l,k} r_{l',k} \big\rangle_k - \big\langle r_{l,k}\big\rangle_k \big\langle r_{l',k}\big\rangle_k }{ a_l\, a_{l'} }
\label{eq:featcorr}
\end{equation}
via the Pearson correlation of their response. 

We find that in the \textit{frozen} region the input weakly activates the features, which are almost perfectly correlated with one another; therefore, the feature graph is almost entirely degenerate. Driving past the onset breaks that degeneracy: duplicates almost vanish, and the activity remains recoverable from the single and two-body sectors that are actually measured. In the \textit{scrambled} region the input is instead dispersed across the network's state faster than what can be recovered from local and two-body features, with activity moving into the high-order sectors while features begin to collapse back onto one another. The optimum therefore lies between the two regions, and it coincides with the onset of scrambling. Note that the networks still perform best at the onset of scrambling even if the readout is extended to higher-order computational-basis features, as shown in Fig.~\ref{fig:scaling_N_k}, leading to increased precision at a higher operating cost.
 
% --- memory ----
\subsubsection*{Scalable nonlinear memory from collective relaxation}
\begin{figure*}[t]
  \centering
  \includegraphics[width=0.98\textwidth]{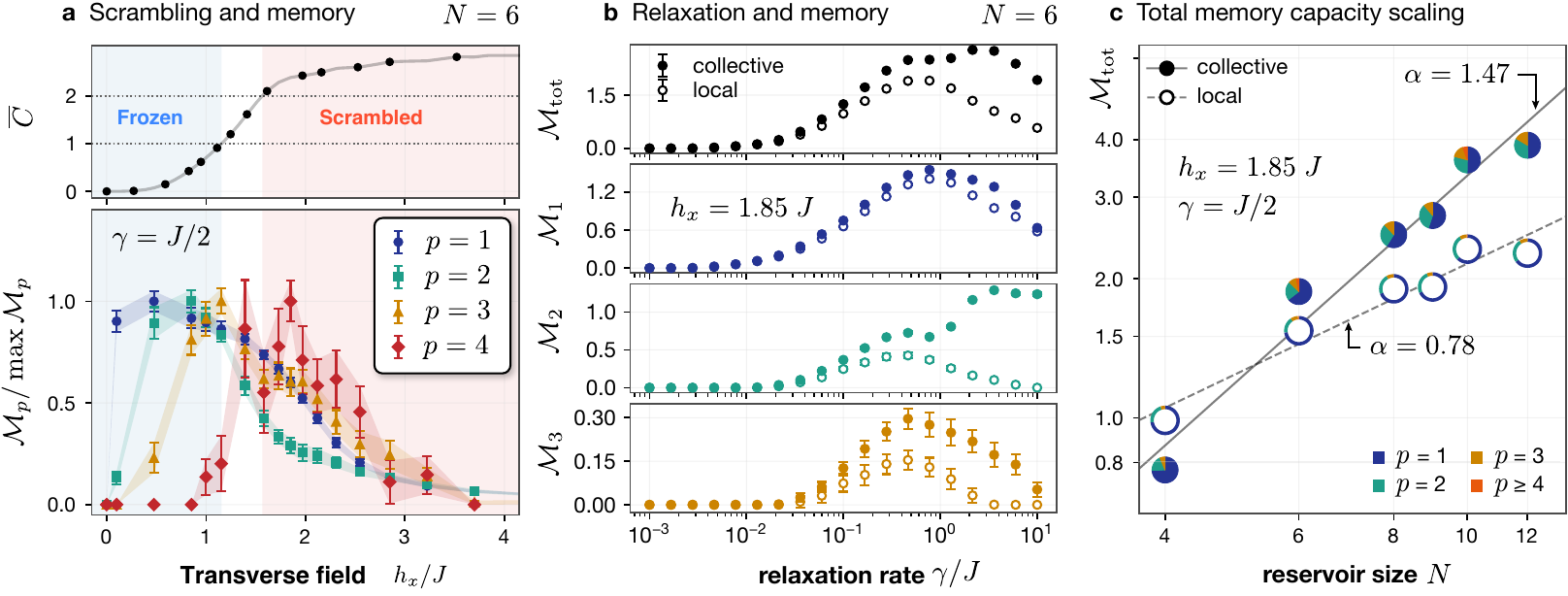}
  \caption{\textbf{Superextensive memory capacity via collective relaxation.}
    \figpanel{a} Memory capacity $\mathcal{M}_p$ of polynomial degree $p = 1$--$4$, each normalised to its own maximum, versus the transverse field for a network of $N = 6$ spins under collective relaxation at rate $\gamma = J/2$. The upper strip shows the OTOC saturation $\overline{C}$ over the same sweep, with the frozen ($\overline{C} < 1$) and scrambled ($\overline{C} > 2$) regions shaded, here obtained by averaging over the time interval $t \in [\pi/2, \pi]$; see Eq.~\eqref{eq:Cbar} for details. Linear memory peaks in the frozen region and successively higher degrees peak at successively stronger fields, so that nonlinear memory requires deeper scrambling than linear memory does.
    \figpanel{b} Total memory capacity $\mathcal{M}_{\mathrm{tot}}$ and its decomposition by degree versus the relaxation rate $\gamma/J$, at $h_x = 1.85\,J$ and $N = 6$, for collective (filled) and local (open) channels. Every contribution is non-monotonic in $\gamma$, with optima at $\gamma/J \approx 1/2$ (local and collective) and $\gamma/J \approx h_x$ (collective). The collective relaxation channel outperforms the local for all $\gamma$ and all $N>4$.
    \figpanel{c} Total memory capacity versus network size at the fixed operating point $h_x = 1.85\,J$, $\gamma = J/2$, with power-law fits $\mathcal{M}_{\mathrm{tot}} \propto N^{\alpha}$ giving $\alpha = 1.47\pm0.15$ for collective and $\alpha = 0.78\pm0.09$ for local relaxation. The two channels cross near $N = 6$, beyond which the collective advantage widens. Markers are shown as pie charts that represent the relative contribution of memory capacity order $\mathcal{M}_p$ as indicated by the legend. Sizes are limited to $N = 12$ by the cost of propagating the open-system dynamics.}
  \label{fig:memory}
\end{figure*}
\noindent
Motivated by feasibility on current hardware, we have so far considered reservoirs that are reset to a fiducial state at every step, with memory supplied by a classical post-processing step. We now relax this condition, allowing the networks to retain their quantum state between inputs, and with it their own memory. This, however, removes the fading memory mechanism required for learning. Indeed, a reservoir that never forgets cannot satisfy the echo-state property~\cite{Boedecker2012}, and its response comes to depend on the entire history of the drive rather than on its recent past. The forgetting must come from somewhere, and in a quantum system the natural source is decoherence~\cite{ChenNurdin2019, Kubota2023,Sannia2024,Kubota2023, Mujal2021}. This setting therefore lets us study how scrambling and decoherence jointly determine what a reservoir remembers.
 
Here, we consider \textit{relaxation} as the fading-memory mechanism, motivated again by availability on current platforms. In particular, we model both \textit{local} $\mathcal{L}_\mathrm{loc}$ and \textit{collective} $\mathcal{L}_\mathrm{col}$ relaxation channels, given in Eqs.~\eqref{eq:local_relaxation} and~\eqref{eq:collective_relaxation} and characterised by a common relaxation rate $\gamma$. In the former, each spin relaxes independently from its excited to its ground state via $\sigma_i^{-}:=\ketbra{0}{1}_i$, so that the network loses its excitations on each site independently. In the latter, the network relaxes through a single jump operator $\sum_i \sigma^-_i$ acting on all spins at once: the decay pathways therefore interfere, and the channel damps only the totally symmetric component of the state, leaving the correlations between spins comparatively intact.

We measure performance by evaluating the information-processing capacity (IPC)~\cite{Dambre2012}, or \textit{memory capacity}, which counts, in a task-independent way, how many independent functions of the input \textit{history} a system can reproduce. This allows us to separate the total memory capacity $\mathcal{M}_\mathrm{tot}$ by its polynomial degree $\mathcal{M}_p$ of the function reproduced, with the linear term ($p=1$) measuring plain memory, higher-order terms ($p>1$) measuring nonlinear memory; see Eq.~\eqref{eq:memory_capacity} in~\methods{} for details.

Sweeping the transverse field at fixed relaxation rate, we find that scrambling does not simply raise or lower the memory of the reservoir, but redistributes it across polynomial orders, as shown in Fig.~\ref{fig:memory}~\figpanel{a}. Linear memory $\mathcal{M}_1$ is largest in the frozen region and decays as scrambling develops. Each successive nonlinear order instead peaks at a progressively stronger field, with the quartic contribution $\mathcal{M}_4$ peaking in the scrambled region. Nonlinearity is therefore acquired with scrambling depth, at the expense of linear memory. As a result, the onset of scrambling is again a sweet spot, where linear and high-order memory contributions are simultaneously appreciable.

Scanning the relaxation rate $\gamma$, we find that memory is non-monotonic for both channels, as shown in Fig.~\ref{fig:memory}~\figpanel{b}. As $\gamma \to 0$ the reservoir retains everything, the echo-state property fails, and the capacity vanishes. At large $\gamma$ the reservoir forgets faster than its dynamics can produce independent features. An optimum lies around $\gamma/J \approx 1/2$ for both local and collective relaxation, where all memory moments are maximal for the local channel. Interestingly, a second optimum lies around $\gamma/J \approx h_x$ for the collective relaxation channel. Crucially, this leads to collective relaxation outperforming local relaxation over the whole range of loss rates $\gamma$, for every order $p$, and at every size $N>4$, with the largest margin in the nonlinear contributions.

That advantage of collective over local relaxation grows with the size $N$ of the network. At a single fixed operating point, $h_x = 1.85\,J$ and $\gamma = J/2$, chosen once and not re-tuned for each size, the total memory capacity grows as $\mathcal{M}_{\mathrm{tot}} \propto N^{\alpha}$ with $\alpha = 1.47$ under collective relaxation and $\alpha = 0.78$ under local relaxation, as shown in Fig.~\ref{fig:memory}~\figpanel{c}. The two channels have comparable memory for $N=4$, crossing near $N = 6$, beyond which the collective advantage grows faster than linearly. Memory therefore joins precision and expressivity in scaling superextensively at the onset of scrambling, provided the reservoir forgets collectively rather than site by site.
 
% --- conclusions ----
\subsection*{Conclusions}
\noindent
In this work, we asked whether information scrambling can serve as a design principle for scalable performance in quantum reservoir computing. Our results indicate that it can. Scrambling, identified through out-of-time-order correlators and quantified against an absolute ergodic reference, organises every performance measure that we have considered: the number of computational-basis states the reservoir populates, its expressive capacity, its forecasting precision, and its memory.

In analogy with edge-of-chaos criticality in classical reservoirs, the onset of scrambling locates the operating region at which the network predicts best. This shows that information scrambling extends the role recently attributed to dynamical phase transitions in quantum reservoirs~\cite{martinezpena2021,Kobayashi2025probe,Kobayashi2026edge}. This picture is consistent with the information-theoretic analysis of Ref.~\cite{Keenan2026}, which appeared while this work was being completed and rests on different foundations. Importantly, the two formalisms converge on identifying scrambling as a key quantity governing reservoir performance. Here we further show how this principle scales with reservoir size towards improved performance and efficiency, while defining an operating \emph{region} that is possibly wider and more robust than expected for a critical point.

This scaling has important consequences on the efficiency of learning. In general-purpose large language models, the loss falls as a small negative power of the number of parameters~\cite{Kaplan2020,Hoffmann2022}, so that precision improves markedly more slowly than the resources devoted to it: at fixed architecture and training recipe, each further increment of quality costs a \textit{larger} amount of computation, and therefore of energy~\cite{Wright2022,Momeni2023,Markovic2020,PNNreview2025}. The reservoirs studied here are task-specific rather than general-purpose, and count physical spins rather than trained parameters, so the comparison sets a general direction rather than a benchmark. Nonetheless, we believe that this direction matters. A superextensive exponent means that performance outruns the resources spent obtaining it, so that the energy cost per unit of precision \emph{falls} as the system is made larger, rather than rising. Whether the scaling persists on hardware is an urgent and exciting question that can be explored on today's devices.

Several avenues can be explored from here. Our results suggest that the reported performance scaling might not be confined to fault-tolerant quantum hardware, since it relies only on a driven, disordered, interacting quantum many-body system with a tunable ratio of coupling to drive, a local encoding, and a low-order readout~\cite{Palacios2024}. Quantum annealers~\cite{Johnson2011}, neutral-atom simulators~\cite{Ebadi2021,Scholl2021}, trapped-ion simulators~\cite{Monroe2021} and coherent photonic Ising machines~\cite{Mohseni2022,Vandoorne2014} all supply these ingredients, and should therefore support superextensive scaling. Encouragingly, superextensive scaling of charging precision has already been observed on a quantum annealer, and survives the decoherence present in the device~\cite{Donelli2025}. Evaluating performance under a finite shot budget is the natural next step, and would open the way to readout strategies that minimise the sampling cost. Likewise, classically efficient paradigms such as tensor-network~\cite{Montangero2018} and neural-network~\cite{Gravina2025, Sinibaldi2026} reservoirs may capture similar behaviour at a polynomial cost in the reservoir size~\cite{Luchnikov2019}.
 
Broadening the benchmark beyond a single classical task and a polynomial readout in the computational basis would also establish the generality of this operating principle. Classification is the natural first extension, having already been demonstrated for quantum reservoirs at scale on image-recognition benchmarks~\cite{Kornjaca2024}, and for the recognition of entangled states and the estimation of nonlinear functionals of a quantum input~\cite{Ghosh2019, TranNakajima2021}. Extending the principle to genuinely quantum inputs, where no classical encoding is required and the encoding bottleneck does not arise, is where a separation from classical reservoirs is most likely to be found, and where the applications of quantum reservoir computing might deliver an advantage over classical computation~\cite{Mujal2021}.

% --- acknowledgments ---
\subsection*{Acknowledgements}
\noindent
This work was supported by the AQSN microgrant awarded by the Hon Hai (Foxconn) Research Institute, and by the Pawsey Supercomputing Research Centre through the National Computational Merit Allocation Scheme 2026. Simulations were carried out within Pawsey's Quantum Supercomputing Innovation Hub, made possible by a grant from the Australian Government through the National Collaborative Research Infrastructure Strategy (NCRIS), using the Setonix supercomputer. F.C. thanks P. Elhai, C. Myers, L. Antoncich, E. Matwiejew and the Pawsey team for insightful discussion and technical support on Setonix. J.F. and F.C. thank V. Baccetti, I. Al-Azki, and J. Cole for inspiration and conversations on criticality in neuromorphic computing and machine learning. F.C. also thanks Q. Tran for discussions on quantum reservoir computing, and W. Lienert, L. Osborne, H. Corless, and B. Tonekaboni for work on related projects. F.C. thanks V.F. for her unwavering support.
 
% --- data ---
\subsection*{Data availability}
\noindent
The data and code that support the findings of this study are available from the corresponding author upon reasonable request.

% --- contributions ----
\subsection*{Author contributions}
\noindent
F.C. conceived the study of information scrambling in quantum Ising networks as a route to scalable performance in quantum reservoir computing for time-series prediction, and developed the framework for its numerical implementation. J.F. contributed to the code for numerical simulations and carried out the systematic evaluation of network performance across regimes and system sizes using local and high-performance computing resources. Both authors discussed and analysed the results, contributing to writing the manuscript.

% --- methods ----
\subsection*{Methods}
\hypertarget{methods}{}

\subsubsection*{Reservoir Hamiltonian}
\noindent
The reservoirs consist of transverse-field quantum Ising networks of $N = L_x \times L_y$ spins on a two-dimensional rectangular lattice, governed by the time-independent mixed-field Ising Hamiltonian of Eq.~\eqref{eq:H}. The couplings carry weak disorder, $J_{ij} = J(1+\delta_{ij})$ with $J > 0$ and $\delta_{ij}$ normally distributed with zero mean and standard deviation $\sigma_J$. The longitudinal field also carries weak normally-distributed disorder, with mean $h_z$ and standard deviation $\sigma_{h_z}$. Disorder in the longitudinal field breaks the $\mathbb{Z}_2$ symmetry of the transverse-field Ising model, making the dynamics non-integrable~\cite{Banuls2011,KimHuse2013}, while disorder in the couplings lifts the degeneracies of the ordered lattice~\cite{Fujii2017,martinezpena2021}. We keep both disorder and mean longitudinal field $h_z<0$ weak relative to $J$, leaving the single dimensionless ratio $h_x/J$ as the main control parameter, while remaining within reach of neutral-atom and superconducting platforms~\cite{Bravo2022,Yasuda2023,Kornjaca2024}. Without loss of generality we set $J = 1$. Throughout this work, we set $\sigma_J = 0.3$, $h_z =- 0.1$, and $\sigma_{h_z} = 0.03$.

\subsubsection*{Encoding, evolution, and readout}
\noindent
The reservoir is initialised in the product state $\ket{\psi_0} = \bigotimes_i \ket{0}_i$, where $\{\ket{0}_i, \ket{1}_i\}$ are the eigenstates of $Z_i$ defining the computational basis. This state is straightforward to prepare on hardware and is far from an eigenstate of $H_0$ at finite $h_x$, so that the subsequent dynamics explore a large portion of the accessible state space. At step $k$, the input value $f_k$---which can be rescaled over its domain---, is written into a fixed subset $\mathcal{I}$ of $N_{\mathrm{in}
} = \lfloor N/2 \rfloor$ spins through an additional transverse drive,
\begin{equation}
H_k = H_0 + h_x f_k \sum_{i \in \mathcal{I}} X_i ,
\label{eq:Hk}
\end{equation}
under which the network evolves for a fixed interval $\tau$, 
\begin{equation}
    \label{eq:dynamics}
    \ket{\psi_k} = U_k \ket{\psi_0}, 
\end{equation}
where
\begin{equation}
    \label{eq:unitary}
    U_k = \exp(-i H_k \tau).
\end{equation}
Here we choose $\tau = \pi/2$. While the results are sensitive to this choice, $\tau$ can be absorbed into the other couplings via a global rescaling of the total Hamiltonian $H_k$, independently of $k$.

The raw features are the expectation values given in Eq.~\eqref{eq:features},
measured over all spins in the network, since all sites are read out in this work. By limiting feature collection to up to $s$-body operators in the computational basis, the total number of features is
\begin{equation}
    \label{eq:featurecount}
    \mathcal{R}_s:=\sum_{s'=1}^{s}\binom{N_\mathrm{out}}{s'},
\end{equation}
where $N_\mathrm{out}$ represent the number of readout spins (here $N_\mathrm{out} = N$).
As a result, there are up to $\mathcal{R}_2 = N(N+1)/2 \sim \mathcal{O}(N^2)$ available features in our framework. 

Because the reservoir is reset after every measurement, $\ket{\psi_k}$ depends on $f_k$ alone, and temporal context is supplied classically. The features passed to the readout are built by an exponentially weighted running average of the raw features, combined with a cyclic permutation of the feature vector,
\begin{equation}
m_{l,k} = \kappa\, m_{\pi(l),\,k-1} \;+\; (1-\kappa)\, r_{l,k} ,
\label{eq:classical_memory}
\end{equation}
with $m_{l,0} = 0$, obtained by adapting the approach introduced in Ref.~\cite{Settino2025}. Here $\kappa \in [0,1)$ is the memory parameter, so that $\kappa = 0$ for raw features and $\kappa > 0$ when classical memory is applied. The index map $\pi(l)$ is the cyclic predecessor of the feature index $l$, so that the retained history is shifted by one feature slot at every step. The input from $j$ steps ago is retained with weight $\kappa^{j}$, and is displaced by $j$ feature slots, so that it does not overwrite more recent inputs. Memory is obtained without enlarging the feature vector, and $\kappa$ alone controls its depth, with a characteristic retention time $1/\ln(1/\kappa)$. 

The readout weights $w_{l}$ are obtained by ridge regression~\cite{Hoerl1970}, with the regularisation strength selected on a validation split, here selected to be 30:70 for testing:training over 800 total steps and a warm up set of 30 steps, and are fitted independently for each forecast horizon. Numerically, this is done using tools from the open sources Python library ReservoirPy~\cite{Trouvain2020}. The classical memory $\kappa\in[0,1]$ is chosen via global optimisation for each selected forecast horizon. 
 
\subsubsection*{Measurement cost}
\noindent
All observables read out here are diagonal in the computational basis and mutually commuting, so a set of projective measurements estimates every feature simultaneously. Each feature is bounded, $|r_{l,k}| \leq 1$, so estimating one of them to additive precision $\eta$ with confidence $1-\delta$ requires $\mathcal{O}(\eta^{-2} \log \delta^{-1})$ shots~\cite{hoeffding1963probability}. A union bound over the $\mathcal{O}(N^{s})$ correlators of weight at most $s$ then gives a total budget of $\mathcal{O}\big(\eta^{-2}(s \log N + \log \delta^{-1})\big)$ shots, so that the sampling cost grows logarithmically with the size of the network while the number of features it delivers grows as $\mathcal{O}(N^{s})$. The cost of estimating all the computational basis correlators up to $s=N$ is therefore linear in $N$.

For the readout used here, $s=2$, a single setting therefore yields $\mathcal{O}(N^{2})$ features at a shot cost essentially independent of $N$. Device-specific accounting, including readout error, the cost of state preparation, and the classical cost of processing $\mathcal{O}(N^{2})$ features, lies beyond the scope of this work.
\begin{figure*}[t]
  \centering
  \includegraphics[width=0.98\textwidth]{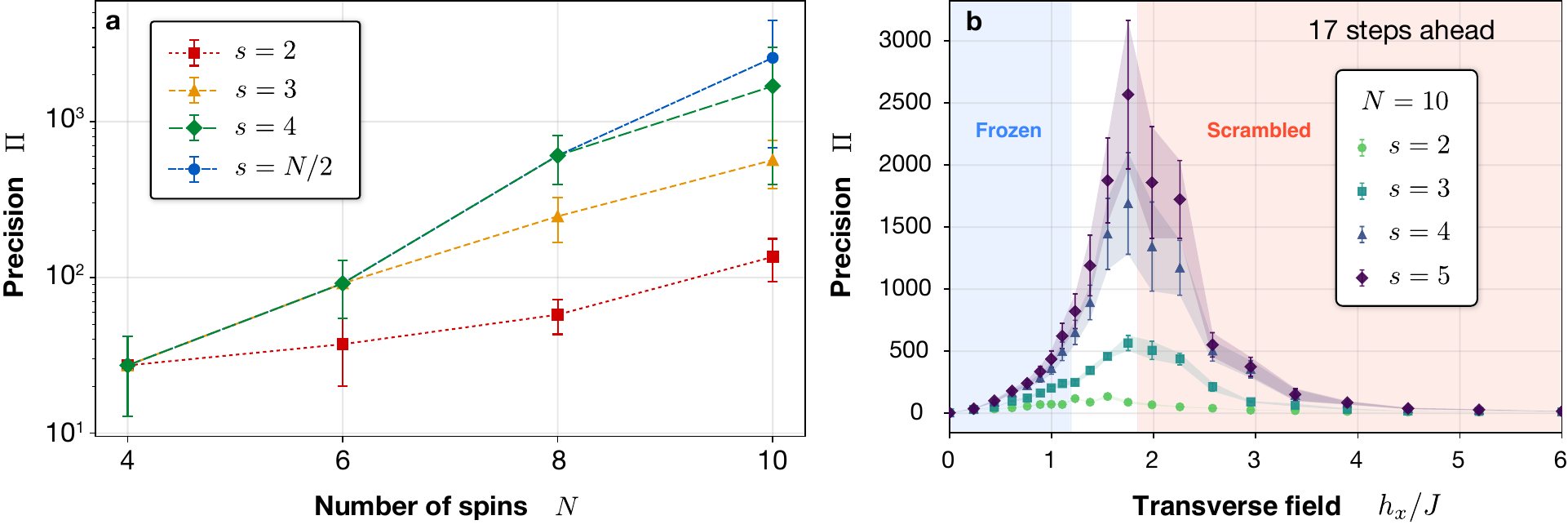}
  \caption{\textbf{Precision scaling with system size under different feature space sizes.} Precision scaling with the number  $N$ of spins and the truncation order $s$ of the features, obtained on the Mackey-Glass forecasting task with a 17 step-ahead prediction horizon, corresponding to the characteristic delay $\tau_\mathrm{MG}=17$. Each truncation retains $\mathcal{R}_s$ features, as per Eq.~\eqref{eq:featurecount}, associated with up to $s$-body correlations. \figpanel{a} Markers denote the highest precision over a 21-point $h_x/J$ sweep; error bars denote $\pm1\sigma$ over 10 realisations. \figpanel{b} Precision across the $h_x/J$ sweep for $N=10$ and $s=2, \dots, 5$, peaking at the onset of scrambling for all values of $s$. Note that for $s=5$ we reach the largest sector of the total feature space of the reservoir, saturating precision, as shown in panel~\figpanel{a} for $s=N/2$. Increasing $s$ beyond $\lceil N/2\rceil$ brings performance into the regime of diminishing returns.}
  \label{fig:scaling_N_k}
\end{figure*}

\subsubsection*{Task and figure of merit}
\noindent
The benchmark that we considered for time-series prediction task is the Mackey--Glass series~\cite{mackey1977oscillation}, generated by the delay differential equation
\begin{equation}
\frac{d f}{d t} = \beta \, \frac{f(t-\tau_{\mathrm{MG}})}{1 + f(t-\tau_{\mathrm{MG}})^{n}} - \gamma_{\mathrm{MG}} \, f(t),
\label{eq:mackeyglass}
\end{equation}
with the standard parameters $\beta = 0.2$, $\gamma_{\mathrm{MG}} = 0.1$, $n = 10$ and delay $\tau_{\mathrm{MG}} = 17$, for which the attractor is chaotic and low-dimensional. The integration step is chosen to be $\Delta t = 1$ so that 17 steps correspond to the intrinsic timescale $\tau_\mathrm{MG}$. The series is generated as in Ref.~\cite{Trouvain2020} and forecast at horizons from one to $85$ steps ahead, as specified case by case.

Performance is quantified by the precision $\Pi = \varepsilon^{-1}$, where
\begin{equation}
\varepsilon = \sqrt{\frac{\sum_k (f_k - \hat{f}_k)^2}{\sum_k (f_k - \bar{f})^2}}
\label{eq:precision}
\end{equation}
is the normalised root-mean-square error (NRMSE), $f_k$ the target, $\hat{f}_k$ the prediction and $\bar{f}$ the mean of the target over the test set. Precision is dimensionless, and larger values indicate better forecasts. Fig.~\ref{fig:scaling_N_k} shows precision scaling as a function of $N$ and of order $s$ of the correlators.

The Mackey-Glass series is generated at $800$ time-steps, with the initial $30$ steps being discarded, and $100$ steps reserved at the end ensuring that a prediction exists at every forecast horizon considered. With forecast horizons spanning $1$ to $85$ steps ahead, of which $17$ (the intrinsic Mackey-Glass timescale $\tau$), $42$ and $84$. A $70:30$ split of the series is dedicated to the training:testing ratio, with each horizon having a tolerance of $10^{-4}$ for ridge regression optimisation. The total number of realisation of network disorder and of input function depends on $N$, with 10 realisations up to $N = 18$ and 5 realisations for $N = 20$.

\subsubsection*{Scrambling diagnostics}
\noindent
The out-of-time-order correlator proper,
\begin{equation}
F_{ij}(t) = \bra{\psi_0} Z_i(t)\, Z_j\, Z_i(t)\, Z_j \ket{\psi_0},
\label{eq:F}
\end{equation}
with $Z_i(t) = U_0^\dagger(t)\, Z_i\, U_0(t)$ and  $U_0(t) = \exp(-i H_0 t)$, begins at unity and decays as information spreads. The squared commutator, given in Eq.~\eqref{eq:otoc}, 
begins at zero and grows. For Hermitian, unitary operators such as the Pauli $Z_i$ the two are equivalent, $C_{ij} = 2\,(1 - \mathrm{Re}\, F_{ij})$, and we work with $C_{ij}$ throughout because it increases with the quantity of interest. It is evaluated on the initial state and under the undriven Hamiltonian, so that the diagnostic is independent of the task~\cite{Fan2017}. The scalar saturation quoted in the main text is a flat double average over the late half of the time grid and over every site other than the source,
\begin{equation}
\overline{C}_i = \frac{1}{(N-1)\,(k_\mathrm{max} - k_0)} \sum_{j\neq i} \sum_{k=k_0}^{k_\mathrm{max}-1} C_{ij}(t_k),
\label{eq:Cbar}
\end{equation}
with $k_0 = \lfloor k_\mathrm{max}/2 \rfloor$, where $k_\mathrm{max}$ is the maximum time-step considered in the OTOC dynamics simulation. Due to the computational cost of evaluating $\overline{C}_i$, here, we only evaluate OTOCs based on a single source site $i = 1$ and drop the index to report $\overline{C}:=\overline{C}_1$. Because $\mathrm{Re}\,F \in [-1,1]$, the squared commutator is bounded, $\overline{C} \in [0,4]$, with three landmarks. At $\overline{C} = 0$ the perturbation and the measurement still commute and no information has travelled. At $\overline{C} = 2$ they are completely decorrelated, $F = 0$, which is the most that scrambling can achieve. The upper limit $\overline{C} = 4$ requires $F = -1$, that is perfect anticorrelation between the perturbed and unperturbed measurements, which is a more ordered situation than a random one and is not reached by scrambling dynamics.

The value $\overline{C} = 2$ is not a convention but the value produced by a completely random evolution, obtained by replacing $U_0(t)$ with a unitary $U$ drawn from the Haar measure, the uniform distribution over unitaries on the $d = 2^N$ states of the network, and averaging. We denote this average by $\mathbb{E}_{\mathrm{Haar}}[\,\cdot\,]$. Since $F_{ij}$ contains the evolved operator $W(t) = U^{\dagger} W U$ twice, the required object is the two-fold twirl $\mathbb{E}_{\mathrm{Haar}}[U^{\dagger}WU \otimes U^{\dagger}WU]$. Averaging over the group leaves only the operators invariant under acting with the same unitary on both copies, of which there are two: the identity $\mathbb{1}$ and the swap $\mathrm{S}$. For traceless Hermitian unitary $W$, with $\mathrm{tr}\,W = 0$ and $\mathrm{tr}\,W^{2} = d$, the coefficients are fixed,
\begin{equation}
\mathbb{E}_{\mathrm{Haar}}\big[ U^{\dagger} W U \otimes U^{\dagger} W U \big]
= \frac{-1}{d^{2}-1}\, \mathbb{1} + \frac{d}{d^{2}-1}\, S .
\label{eq:twirl}
\end{equation}

Contracting Eq.~\eqref{eq:twirl} with two copies of the probe operator $V$, using $\mathrm{tr}[W(t) V W(t) V] = \mathrm{tr}[(W(t) \otimes W(t))(V \otimes V)\mathrm{S}]$ together with $\mathrm{tr}[(V \otimes V)\mathrm{S}] = \mathrm{tr}\,V^{2} = d$ and $\mathrm{tr}[\mathrm{S}(V \otimes V)\mathrm{S}] = (\mathrm{tr}\,V)^{2} = 0$, gives
\begin{equation}
\mathbb{E}_{\mathrm{Haar}}[F] = \frac{-1}{d^{2}-1} ,
\qquad
\mathbb{E}_{\mathrm{Haar}}\big[\overline{C}\big] = 2 + \frac{2}{d^{2}-1} \simeq 2 .
\label{eq:haar}
\end{equation}

The residual is exponentially small in the number of spins, so a fully scrambled network is indistinguishable from a random one at the level of this observable, and $\overline{C} = 2$ serves as an absolute reference rather than a fitted one. The twirl involves only two copies of $U$, so any ensemble reproducing the correct two-copy average---a two-design, such as the classically simulable Clifford group---yields the same value. Reaching $\overline{C} = 2$ therefore certifies scrambling but not chaos~\cite{Hosur2016,RobertsYoshida2017,Dowling2023}.

\subsubsection*{Resolvable expressive capacity and bounds}
\noindent
Here, we evaluate the expressive capacity $\mathcal{C}$ by calculating the rank of the Gram matrix $\mathbf{G}:=\mathbf{m} \:\mathbf{m}^T$ of the feature matrix $\mathbf{m}$, whose elements are given by the features $m_{l,k}$, evaluated in the infinite-shot noiseless limit,
\begin{equation}
    \label{eq:expressive_capacity}
    \mathcal{C} = \mathrm{Rank}\{\mathbf{G}\}.
\end{equation}
To evaluate the rank, we calculate the number of non-zero eigenvalues $\lambda_k$ of $\mathbf{G}$, using a tolerance of $\lambda_k/\max{k}\{\lambda_k\} > 10^{-9}$.

When features are estimated based on a shot budget, the resolvable expressive capacity (REC), $\mathcal{C}_S$, quantifies how many linearly independent functions a physical system expresses from its outputs under $S$ measurement shots~\cite{Hu2023REC},
\begin{equation}
\mathcal{C}_S
=
\sum_{k=0}^{K-1}
\frac{1}{1+\beta_k^2/S},
\label{eq:shot-based-rec}
\end{equation}
where $k\in\{0,\dots,K-1\}$ spans the total number of measured degrees of freedom of a system (here, the raw features), with $K$ being the maximum amount of features, and where $\beta_k^2$ is the number of shots that must be spent to resolve the signal from the $k$-th feature~\cite{Hu2023REC}. Note that $\beta_k^2$ can be obtained from the generalised eigenvalue problem to the estimated Gram matrix and the input-averaged sampling covariance matrix~\cite{Hu2023REC}. As a result, $S/\beta_k^2$ corresponds to the signal-to-noise ratio needed to resolve the $k$-th feature. In the infinite-shot limit $S\to\infty$ the noise contribution vanishes and the REC reduces to
the rank of the Gram matrix of Eq.~\eqref{eq:expressive_capacity},
\begin{equation}
    \label{eq:infinit_shot_limit_rec}
    \mathcal{C} = \lim_{S\to\infty}\mathcal{C}_S.
\end{equation}

In digital quantum reservoir computing with gate-based encoding the input is written by unitary gates and each readout is a truncated multivariate
Fourier series in the input whose frequencies are fixed by the eigenvalue
differences encoding generator~\cite{Schutte2025, holzer2026spectral}. For $N_\mathrm{in}$ encoding qubits this gives a total number of non-zero eigenvalues $\Omega$
\begin{equation}
\Omega =
\begin{cases}
2N_\mathrm{in}+1, & \text{Hamming } \\[2pt]
3^{N_\mathrm{in}}, & \text{Ternary } \\[2pt]
4^{N_\mathrm{in}}-2^{N_\mathrm{in}}+1, & \text{Golomb } 
\end{cases}
\label{eq:encoding-hierarchy}
\end{equation}
where the \textit{Hamming} case is the common-weight single-qubit encoding used here, the \textit{ternary} case assigns incommensurate weights in geometric progression, and the \textit{Golomb} case uses multi-qubit generators whose eigenvalues form a Sidon set~\cite{sidon1932satz}, which is the largest spectrum attainable on $N_\mathrm{in}$ input qubits~\cite{holzer2026spectral}. 
\begin{figure}[t]
  \centering
  \includegraphics[width=0.48\textwidth]{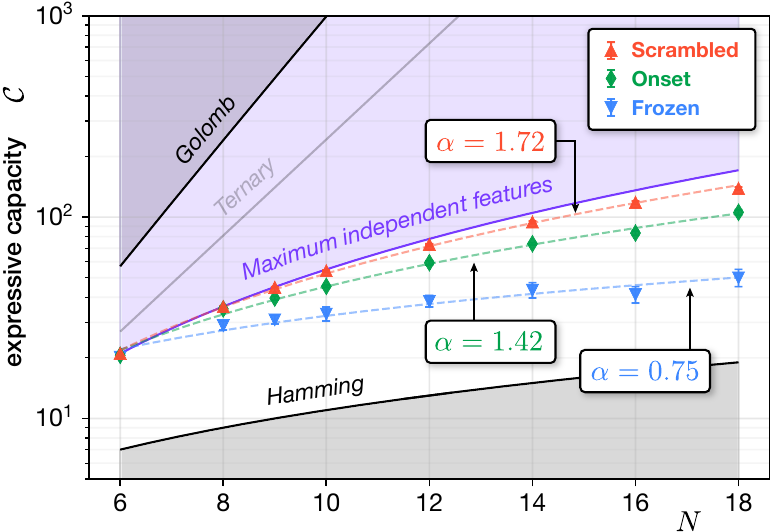}
  \caption{\textbf{Scrambling enables superextensive expressivity.}
  Expressive capacity, measured as the number of linearly independent directions present in the feature set, versus network size $N$, evaluated in the frozen, onset, and scrambled regions. Dashed lines are power-law fits $\propto N^\alpha$, giving $\alpha = 0.75\pm0.07$ when frozen, $\alpha = 1.42\pm0.04$ at the onset and $\alpha = 1.72\pm0.03$ when scrambled. 
  Also shown are some notable bounds represented with solid lines. The maximum number of independent features supported by local and two-body readout (purple), the ternary and Golomb encoding bounds (grey and black, upper shaded region), which grow exponentially in $N$, and the Hamming bound (lower shaded region), linear in $N$. Expressive capacity is evaluated in the infinite-shot noiseless limit. Error bars denote the standard error over disorder realisations.}
  \label{fig:rec}
\end{figure}

Our encoding is analog rather than digital, gate-based. As a result, the input evolves with the reservoir dynamics, where the encoding Hamiltonian $H_1(t_k)$, local, does not in general commute with the bare Hamiltonian of the reservoir $H_0$, interacting, $[H_0, H_1(t_k)]\neq0$. Our features are also mixed via the classical memory step of Eq.~\eqref{eq:classical_memory}, so these limits do not apply, as shown in Fig.~\ref{fig:rec}. The infinite-shot noiseless EC is therefore bounded simultaneously by encoding and readout,
\begin{equation}
\mathcal{C} = \operatorname{Rank}\{\mathbf{G}\} \leq \min(\Omega,\,\mathcal{R}),
\label{eq:min-bound}
\end{equation}
where $\mathcal{R}$ is the total number of features; see Eq.~\eqref{eq:featurecount}.

\subsubsection*{Memory capacity and information processing capacity}
\label{sec:MC}

\noindent
We quantify what the reservoir remembers, linearly and nonlinearly, through the information processing capacity (IPC)~\cite{Dambre2012,MartinezPena2023IPC}. Rather than measuring performance on one chosen task, the IPC asks the reservoir to reproduce a large family of \textit{target functions} of its own input history, and counts how many of them it can reconstruct. In this section the reservoir retains its quantum state between inputs, so its memory is its own: the readout is therefore trained on the raw features $r_{l,k}$ of Eq.~\eqref{eq:features}, without the classical memory step of Eq.~\eqref{eq:classical_memory}, and the classical memory parameter is set to $\kappa = 0$ throughout.

We drive the reservoir with an input $f_k$ drawn independently at each step from a uniform distribution on $[-1,1]$, rather than with the Mackey--Glass series used for forecasting, since the IPC requires an input with no temporal correlations of its own. A \textit{target} is any function $y$ of the past inputs, whose value at step $k$ we write $y_k$. The reservoir's reconstruction of it is
\begin{equation}
\hat{y}_k = w_0 + \sum_{l} w_l\, r_{l,k},
\label{eq:ipc-readout}
\end{equation}
a ridge readout of the raw features with a fitted bias $w_0$, and the quality of that reconstruction is the coefficient of determination
\begin{equation}
\chi[y] = \max\!\left(0,\; 1 - \frac{\sum_k\left(y_k - \hat{y}_k\right)^{2}}{\sum_k\left(y_k - \bar{y}\right)^{2}}\right),
\label{eq:capacity}
\end{equation}
with $\bar{y}$ the mean of the target over the training steps. We call $\chi[y] \in [0,1]$ the capacity for target $y$: it equals $1$ for exact reconstruction and $0$ when the reservoir does no better than predicting the mean. Weights are fitted on a training split and Eq.~\eqref{eq:capacity} is evaluated on the held-out remainder.

Targets are built from delayed inputs. The simplest are the delayed inputs themselves, $y_k = f_{k-\Delta}$, where $\Delta$ represents a delay, and summing their capacities over delays gives the \textit{linear} memory capacity,
\begin{equation}
\mathcal{M}_1 = \sum_{\Delta=1}^{\Delta_{\max}} \chi\!\left[f_{k-\Delta}\right]\Theta\!\left(\chi[f_{k-\Delta}] - \chi_0\right),
\label{eq:mc}
\end{equation}
which is the quantity introduced by Jaeger~\cite{jaeger2001echo} and measures plain recall of past inputs. Unlike the original formulation, we include the Heaviside step function $\Theta$, as in Dambre~\cite{Dambre2012}, to suppress capacities below the noise floor $\chi_0$, since at finite series length a small nonzero $\chi$ can arise from statistical fluctuations alone. Note that we choose a maximum delay $\Delta_\mathrm{max}=\min(20,2N)$ that depends on $N$ because the memory kernel extends within larger reservoirs. An $N$-spin reservoir loses its memory much earlier than a $2N$-spin reservoir. Our choice is empirical and based on convergence of the results by imposing larger cut-offs on the maximum delay size. 

Nonlinear targets are products of Legendre polynomials $\tilde{P}_{p_\nu}$ evaluated at distinct past inputs,
\begin{equation}
y_k = \prod_{\nu=1}^{n} \tilde{P}_{p_\nu}\!\left(f_{k-\Delta_\nu}\right),
\qquad \sum_{\nu=1}^{n} p_\nu = p,
\label{eq:ipc-set}
\end{equation}
where $n$ is the number of distinct past times entering the target, $\Delta_1 < \dots < \Delta_n$ are their delays, and $p_\nu \geq 1$ are the degrees assigned to each. The total degree $p = \sum_\nu p_\nu$ labels the target: $p = 1$ recovers the delayed inputs of Eq.~\eqref{eq:mc}, $p = 2$ contains both squares of a single past input and products of two different ones, and so on. Normalised Legendre polynomials are used because they are mutually orthogonal under the uniform input measure, so that the capacities of different targets are additive and cannot double-count the same reconstructed function.

Writing $\mathcal{Y}_p$ for the set of all targets of total degree $p$, the memory capacity at that degree is
\begin{equation}
\mathcal{M}_p = \sum_{y \in \mathcal{Y}_p} \chi[y]\; \Theta\!\left(\chi[y] - \chi_0\right),
\label{eq:Mp}
\end{equation}
from here, the \textit{total} memory capacity quoted is the sum over degrees, 
\begin{equation}
    \label{eq:memory_capacity}
    \mathcal{M}_\mathrm{tot}=\sum_{p=1}^{p_{\max}}\mathcal{M}_p,
\end{equation}
where $\mathcal{M}_p$ is the capacity at degree $p$ and $p_{\max}$ the highest degree evaluated, and is bounded above by the number of linearly independent readout features.

We use $\Delta^{\mathrm{nl}}_{\max} = 8$ for the nonlinear delay window, at most $n_{\max} = 3$ distinct past times per target, degrees up to $p_{\max} = 4$, and $\chi_0 = 0.03$. These choices give $|\mathcal{Y}_2| = 36$, $|\mathcal{Y}_3| = 92$ and $|\mathcal{Y}_4| = 120$ targets, where here $|\cdot|$ represents the number of elements in a set. The nonlinear window is shorter than the linear one because $|\mathcal{Y}_p|$ grows combinatorially with it: widening it to $\Delta^{\mathrm{nl}}_{\max} = 20$ would raise $|\mathcal{Y}_4|$ from $120$ to $1540$. For the same reason the enumeration is capped at $60$ targets per degree, so that $\mathcal{M}_2$ is scored in full while $\mathcal{M}_p$ for $p \geq 3$ is a lower bound obtained from the first $60$ elements of $\mathcal{Y}_p$ in a fixed ordering, identical for all $\gamma$ and $N$.

\subsubsection*{Open-system dynamics}
\noindent
Open-system dynamics are simulated via the Lindblad master
equation~\cite{lindblad1976}, which governs the time evolution of the density
matrix $\rho$ of the reservoir,
\begin{equation}
    \dot{\rho} = -i\big[H(t), \rho\big] +\mathcal{L}[\rho],
    \label{eq:lindblad}
\end{equation}
where $\mathcal{L}$ denotes the relaxation channel. We compare two channels at a rate
$\gamma$: \emph{local} relaxation through $N$ independent operators,
\begin{equation}
    \label{eq:local_relaxation}
    \mathcal{L}_\mathrm{loc}[\rho]:=\sum_i \left( R_i \rho R_i^{\dagger}
               - \frac{1}{2}\big\{ R_i^{\dagger} R_i, \rho \big\} \right),
\end{equation}
with $R_i = \sqrt{\gamma}\,\sigma^-_i,$, where $\sigma_i^- = \ketbra{0}{1}_i$, in which each spin decays independently and $\{\cdot,\cdot\}$ denotes the anticommutator, and \emph{collective} relaxation through the single operator
\begin{equation}
    \label{eq:collective_relaxation}
        \mathcal{L}_\mathrm{col}[\rho]:=\left( R \rho R^{\dagger}
               - \frac{1}{2}\big\{ R^{\dagger} R, \rho \big\} \right),
\end{equation} 
where $R = \sqrt{\gamma}\sum_i \sigma^-_i$. In this case the state persists between input steps and is not reset to the fiducial state $\ket{\psi_0}$ unlike in the closed-system dynamics. 

Propagating Eq.~(\ref{eq:lindblad}) directly costs $\mathcal{O}(4^N)$ in memory, which caps exact density matrix integration near $N \approx 10$. We therefore use the Monte Carlo wavefunction method, i.e., quantum trajectories, and integrate with QuTiP~5.3.0's \texttt{mcsolve}, to ensure the state remains a $2^N$ vector and $N = 12$ remains tractable; predictions are averaged over $100$ trajectories per realisation with $10$ realisations for $N=4,6,8$ and $2$ realisations at $N=10,12$. Correctness is verified at small $N$ by comparing the trajectory average against exact \texttt{mesolve} integration of the full density matrix. 

\subsubsection*{Numerical simulation}
\begin{figure}[t]
  \centering
  \includegraphics[width=0.48\textwidth]{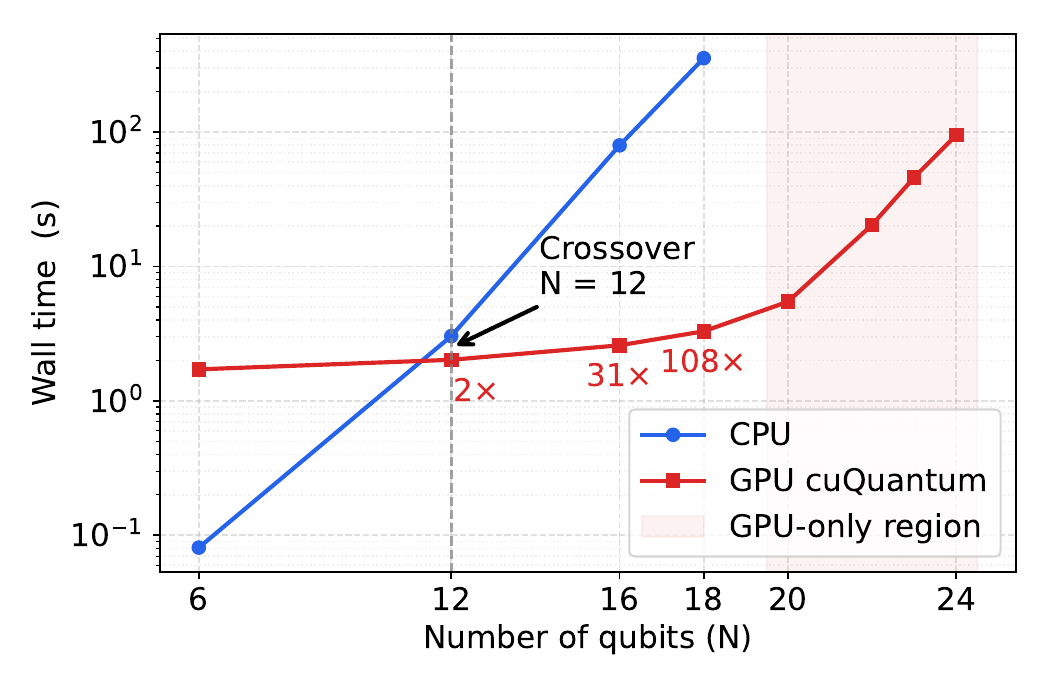}
  \caption{\textbf{CPU-GPU Crossover in QRC Simulation Wall-Time.} Wall time, expressed in seconds ($\mathrm{s}$) per QRC trajectory as a function of system size ($N$) on the Setonix-Q GH200 superchip, comparing the 72-core Grace ARM CPU using QuTiP (blue), with the H100 GPU using the cuStateVec Trotter backend (red). The CPU cost grows exponentially at $2^{N}$, while
GPU time is contained below $10$ seconds up to $N \approx 20$. The curves cross at $N=12$. Beyond $N \approx 18$, GPU wall time also exhibits exponential scaling, reaching $\approx 100$ seconds at  $N=24$. Red shading denotes system sizes  $N=20,22,23,24$ evaluated only on the GPU.}
  \label{fig:Pawsey}
\end{figure}

\noindent
Two independent computational backends were used. CPU was used for exact propagation, integrating the Schr\"odinger equation directly with QuTiP~5.3.0's \texttt{sesolve} on a Hamiltonian. This approach is free of approximation errors, but limited to $N \lesssim 18$ by wall-time cost (Fig.~\ref{fig:Pawsey}). GPU was used for trotterised cuStateVec propagation. The GPU path uses NVIDIA cuQuantum/cuStateVec~1.13.1~\cite{bayraktar2023cuquantum} through \texttt{cuquantum-python-cu12}~26.3.2. Each step resets the CuPy state vector to the product state and applies a second-order Trotter-Suzuki decomposition~\cite{suzuki1976relationship} of the propagation unitary $U_k = e^{-iH_k\tau}$ in $m$ trotter-steps,
\begin{equation}
  U_k=
  \Big[e^{-iH_X \delta/2}\; e^{-iH_Z \delta}\; e^{-iH_X \delta/2}\Big]^{m}
  +\mathcal{O}(\tau\delta^{2}),
  \label{eq:strang}
\end{equation}
where $\mathcal{O}(\tau\delta^2)$ denotes terms of the order of $\tau\delta^2$, and where $H_X$ collects the transverse terms, including the input modulation, and $H_Z$ the coupling and longitudinal field. Production runs use $m = 128$ trotter-steps at interval $\tau = \pi/2$, verified against the CPU reference in Table~\ref{tab:trotter}.

Simulations were executed on the Pawsey Supercomputing Research Centre
\textsc{Setonix-Q} partition, on NVIDIA GH200 Grace-Hopper superchips, each pairing a Grace CPU with a $96$ GB H100 GPU~\cite{pawsey_setonixq_quickstart}. At the largest size studied for stress test ($N=24$) the run occupies $1.073$ GB of memory, the accessible range is bounded by wall-clock allocation rather than device memory.
 
Convergence of the feature dynamics from Trotterisation was verified against the exact {QuTiP} \texttt{sesolve} method.  At $N=12$ the feature matrix deviates by $\max|\Delta| = 4.0\times10^{-4}$ at $m=128$ trotter-steps, and the deviation decreases by $\mathcal{O}(m^{-2})$ over $m \in [24,256]$ (Table~\ref{tab:trotter}), confirming the expected $(\tau/m)^{2}$ scaling of the Strang splitting \cite{suzuki1976relationship}. At fixed Trotter-steps $m=128$ the deviation is essentially independent of system size across $N=6$-$18$, with the maximum feature matrix deviation being; $2.1$-$3.2\times10^{-4}$, with $N=18$ being the largest size where the CPU backend remains tractable. For both backends the NRMSE agrees to four significant figures at every $N$, both figures sit an order of magnitude inside the $10^{-3}$ tolerance applied before any production job is submitted.

Runs are submitted as \textsc{Slurm} array jobs, one task per configuration, with results stored individually to ensure data is not lost after a job failure. Each realisation is seeded deterministically from its configuration, to ensure reproducibility.

\begin{table}[tb]
\begin{ruledtabular}
\begin{tabular}{rll}
\multicolumn{3}{l}{\figpanel{a} \textbf{Trotter-steps convergence at $N=12$}} \\[2pt]
$m$ & $\tau/m$ & $\max|\Delta|$ \\ \hline
 24 & $6.5\times10^{-2}$ & $1.21\times10^{-2}$ \\
 48 & $3.3\times10^{-2}$ & $2.95\times10^{-3}$ \\
 96 & $1.6\times10^{-2}$ & $7.25\times10^{-4}$ \\
128 & $1.2\times10^{-2}$ & $4.03\times10^{-4}$ \\
192 & $8.2\times10^{-3}$ & $1.77\times10^{-4}$ \\
256 & $6.1\times10^{-3}$ & $1.00\times10^{-4}$ \\
\end{tabular}
\vspace{6pt}
\begin{tabular}{rlll}
\multicolumn{4}{l}{\figpanel{b} \textbf{System size dependence at $m=128$}} \\[2pt]
$N$ & $\max|\Delta|$ & NRMSE (CPU) & NRMSE (GPU) \\ \hline
 6 & $2.08\times10^{-4}$ & $0.01677$ & $0.01674$ \\
12 & $2.96\times10^{-4}$ & $0.02921$ & $0.02922$ \\
16 & $3.19\times10^{-4}$ & $0.01373$ & $0.01373$ \\
18 & $2.95\times10^{-4}$ & $0.01768$ & $0.01767$ \\
\end{tabular}
\end{ruledtabular}
\caption{\textbf{Verification of the Strang-Trotter feature dynamics against
the exact CPU reference.} \figpanel{a} Maximum absolute deviation of the feature matrix
from {QuTiP} \texttt{sesolve} at $N=12$ as a function of the number of
trotter-steps $m$ per input interval $\tau = \pi/2$; each doubling of $m$
reduces the deviation by a factor of $\approx 4$, and a power-law fit gives $\max|\Delta| \propto
m^{-2.03}$. \figpanel{b} Deviation at the production value $m=128$ as a function of
system size, together with the forecasting NRMSE obtained on each backend; All entries lie an order of magnitude inside the $10^{-3}$ acceptance gate.}
\label{tab:trotter}
\end{table}

% --- bibliography ---
\bibliographystyle{apsrev4-2}
\bibliography{main}

\begin{thebibliography}{93}%
\makeatletter
\providecommand \@ifxundefined [1]{%
 \@ifx{#1\undefined}
}%
\providecommand \@ifnum [1]{%
 \ifnum #1\expandafter \@firstoftwo
 \else \expandafter \@secondoftwo
 \fi
}%
\providecommand \@ifx [1]{%
 \ifx #1\expandafter \@firstoftwo
 \else \expandafter \@secondoftwo
 \fi
}%
\providecommand \natexlab [1]{#1}%
\providecommand \enquote  [1]{``#1''}%
\providecommand \bibnamefont  [1]{#1}%
\providecommand \bibfnamefont [1]{#1}%
\providecommand \citenamefont [1]{#1}%
\providecommand \href@noop [0]{\@secondoftwo}%
\providecommand \href [0]{\begingroup \@sanitize@url \@href}%
\providecommand \@href[1]{\@@startlink{#1}\@@href}%
\providecommand \@@href[1]{\endgroup#1\@@endlink}%
\providecommand \@sanitize@url [0]{\catcode `\\12\catcode `\$12\catcode `\&12\catcode `\#12\catcode `\^12\catcode `\_12\catcode `\%12\relax}%
\providecommand \@@startlink[1]{}%
\providecommand \@@endlink[0]{}%
\providecommand \url  [0]{\begingroup\@sanitize@url \@url }%
\providecommand \@url [1]{\endgroup\@href {#1}{\urlprefix }}%
\providecommand \urlprefix  [0]{URL }%
\providecommand \Eprint [0]{\href }%
\providecommand \doibase [0]{https://doi.org/}%
\providecommand \selectlanguage [0]{\@gobble}%
\providecommand \bibinfo  [0]{\@secondoftwo}%
\providecommand \bibfield  [0]{\@secondoftwo}%
\providecommand \translation [1]{[#1]}%
\providecommand \BibitemOpen [0]{}%
\providecommand \bibitemStop [0]{}%
\providecommand \bibitemNoStop [0]{.\EOS\space}%
\providecommand \EOS [0]{\spacefactor3000\relax}%
\providecommand \BibitemShut  [1]{\csname bibitem#1\endcsname}%
\let\auto@bib@innerbib\@empty
%</preamble>
\bibitem [{\citenamefont {Beggs}\ and\ \citenamefont {Plenz}(2003)}]{Beggs2003}%
  \BibitemOpen
  \bibfield  {author} {\bibinfo {author} {\bibfnamefont {J.~M.}\ \bibnamefont {Beggs}}\ and\ \bibinfo {author} {\bibfnamefont {D.}~\bibnamefont {Plenz}},\ }\href {https://doi.org/10.1523/JNEUROSCI.23-35-11167.2003} {\bibfield  {journal} {\bibinfo  {journal} {The Journal of Neuroscience}\ }\textbf {\bibinfo {volume} {23}},\ \bibinfo {pages} {11167} (\bibinfo {year} {2003})}\BibitemShut {NoStop}%
\bibitem [{\citenamefont {Shew}\ \emph {et~al.}(2009)\citenamefont {Shew}, \citenamefont {Yang}, \citenamefont {Petermann}, \citenamefont {Roy},\ and\ \citenamefont {Plenz}}]{Shew2009}%
  \BibitemOpen
  \bibfield  {author} {\bibinfo {author} {\bibfnamefont {W.~L.}\ \bibnamefont {Shew}}, \bibinfo {author} {\bibfnamefont {H.}~\bibnamefont {Yang}}, \bibinfo {author} {\bibfnamefont {T.}~\bibnamefont {Petermann}}, \bibinfo {author} {\bibfnamefont {R.}~\bibnamefont {Roy}},\ and\ \bibinfo {author} {\bibfnamefont {D.}~\bibnamefont {Plenz}},\ }\href {https://doi.org/10.1523/JNEUROSCI.3864-09.2009} {\bibfield  {journal} {\bibinfo  {journal} {The Journal of Neuroscience}\ }\textbf {\bibinfo {volume} {29}},\ \bibinfo {pages} {15595} (\bibinfo {year} {2009})}\BibitemShut {NoStop}%
\bibitem [{\citenamefont {Mu{\~n}oz}(2018)}]{Munoz2018}%
  \BibitemOpen
  \bibfield  {author} {\bibinfo {author} {\bibfnamefont {M.~A.}\ \bibnamefont {Mu{\~n}oz}},\ }\href {https://doi.org/10.1103/RevModPhys.90.031001} {\bibfield  {journal} {\bibinfo  {journal} {Reviews of Modern Physics}\ }\textbf {\bibinfo {volume} {90}},\ \bibinfo {pages} {031001} (\bibinfo {year} {2018})}\BibitemShut {NoStop}%
\bibitem [{\citenamefont {Stieg}\ \emph {et~al.}(2012)\citenamefont {Stieg}, \citenamefont {Avizienis}, \citenamefont {Sillin}, \citenamefont {Martin-Olmos}, \citenamefont {Aono},\ and\ \citenamefont {Gimzewski}}]{Stieg2012}%
  \BibitemOpen
  \bibfield  {author} {\bibinfo {author} {\bibfnamefont {A.~Z.}\ \bibnamefont {Stieg}}, \bibinfo {author} {\bibfnamefont {A.~V.}\ \bibnamefont {Avizienis}}, \bibinfo {author} {\bibfnamefont {H.~O.}\ \bibnamefont {Sillin}}, \bibinfo {author} {\bibfnamefont {C.}~\bibnamefont {Martin-Olmos}}, \bibinfo {author} {\bibfnamefont {M.}~\bibnamefont {Aono}},\ and\ \bibinfo {author} {\bibfnamefont {J.~K.}\ \bibnamefont {Gimzewski}},\ }\href {https://doi.org/10.1002/adma.201103053} {\bibfield  {journal} {\bibinfo  {journal} {Advanced Materials}\ }\textbf {\bibinfo {volume} {24}},\ \bibinfo {pages} {286} (\bibinfo {year} {2012})}\BibitemShut {NoStop}%
\bibitem [{\citenamefont {Hochstetter}\ \emph {et~al.}(2021)\citenamefont {Hochstetter}, \citenamefont {Zhu}, \citenamefont {Loeffler}, \citenamefont {Diaz-Alvarez}, \citenamefont {Nakayama},\ and\ \citenamefont {Kuncic}}]{Hochstetter2021}%
  \BibitemOpen
  \bibfield  {author} {\bibinfo {author} {\bibfnamefont {J.}~\bibnamefont {Hochstetter}}, \bibinfo {author} {\bibfnamefont {R.}~\bibnamefont {Zhu}}, \bibinfo {author} {\bibfnamefont {A.}~\bibnamefont {Loeffler}}, \bibinfo {author} {\bibfnamefont {A.}~\bibnamefont {Diaz-Alvarez}}, \bibinfo {author} {\bibfnamefont {T.}~\bibnamefont {Nakayama}},\ and\ \bibinfo {author} {\bibfnamefont {Z.}~\bibnamefont {Kuncic}},\ }\href {https://doi.org/10.1038/s41467-021-24260-z} {\bibfield  {journal} {\bibinfo  {journal} {Nature Communications}\ }\textbf {\bibinfo {volume} {12}},\ \bibinfo {pages} {4008} (\bibinfo {year} {2021})}\BibitemShut {NoStop}%
\bibitem [{\citenamefont {Tanaka}\ \emph {et~al.}(2019)\citenamefont {Tanaka}, \citenamefont {Yamane}, \citenamefont {H{\'e}roux}, \citenamefont {Nakane}, \citenamefont {Kanazawa}, \citenamefont {Takeda}, \citenamefont {Numata}, \citenamefont {Nakano},\ and\ \citenamefont {Hirose}}]{Tanaka2019}%
  \BibitemOpen
  \bibfield  {author} {\bibinfo {author} {\bibfnamefont {G.}~\bibnamefont {Tanaka}}, \bibinfo {author} {\bibfnamefont {T.}~\bibnamefont {Yamane}}, \bibinfo {author} {\bibfnamefont {J.~B.}\ \bibnamefont {H{\'e}roux}}, \bibinfo {author} {\bibfnamefont {R.}~\bibnamefont {Nakane}}, \bibinfo {author} {\bibfnamefont {N.}~\bibnamefont {Kanazawa}}, \bibinfo {author} {\bibfnamefont {S.}~\bibnamefont {Takeda}}, \bibinfo {author} {\bibfnamefont {H.}~\bibnamefont {Numata}}, \bibinfo {author} {\bibfnamefont {D.}~\bibnamefont {Nakano}},\ and\ \bibinfo {author} {\bibfnamefont {A.}~\bibnamefont {Hirose}},\ }\href {https://doi.org/10.1016/j.neunet.2019.03.005} {\bibfield  {journal} {\bibinfo  {journal} {Neural Networks}\ }\textbf {\bibinfo {volume} {115}},\ \bibinfo {pages} {100} (\bibinfo {year} {2019})}\BibitemShut {NoStop}%
\bibitem [{\citenamefont {Langton}(1990)}]{langton1990}%
  \BibitemOpen
  \bibfield  {author} {\bibinfo {author} {\bibfnamefont {C.~G.}\ \bibnamefont {Langton}},\ }\href {https://doi.org/10.1016/0167-2789(90)90064-V} {\bibfield  {journal} {\bibinfo  {journal} {Physica D: Nonlinear Phenomena}\ }\textbf {\bibinfo {volume} {42}},\ \bibinfo {pages} {12} (\bibinfo {year} {1990})}\BibitemShut {NoStop}%
\bibitem [{\citenamefont {Bertschinger}\ and\ \citenamefont {Natschl{\"a}ger}(2004)}]{Bertschinger2004}%
  \BibitemOpen
  \bibfield  {author} {\bibinfo {author} {\bibfnamefont {N.}~\bibnamefont {Bertschinger}}\ and\ \bibinfo {author} {\bibfnamefont {T.}~\bibnamefont {Natschl{\"a}ger}},\ }\href {https://doi.org/10.1162/089976604323057443} {\bibfield  {journal} {\bibinfo  {journal} {Neural Computation}\ }\textbf {\bibinfo {volume} {16}},\ \bibinfo {pages} {1413} (\bibinfo {year} {2004})}\BibitemShut {NoStop}%
\bibitem [{\citenamefont {Luko{\v{s}}evi{\v{c}}ius}\ and\ \citenamefont {Jaeger}(2009)}]{Lukosevicius2009}%
  \BibitemOpen
  \bibfield  {author} {\bibinfo {author} {\bibfnamefont {M.}~\bibnamefont {Luko{\v{s}}evi{\v{c}}ius}}\ and\ \bibinfo {author} {\bibfnamefont {H.}~\bibnamefont {Jaeger}},\ }\href {https://doi.org/10.1016/j.cosrev.2009.03.005} {\bibfield  {journal} {\bibinfo  {journal} {Computer Science Review}\ }\textbf {\bibinfo {volume} {3}},\ \bibinfo {pages} {127} (\bibinfo {year} {2009})}\BibitemShut {NoStop}%
\bibitem [{\citenamefont {Chialvo}(2010)}]{Chialvo2010}%
  \BibitemOpen
  \bibfield  {author} {\bibinfo {author} {\bibfnamefont {D.~R.}\ \bibnamefont {Chialvo}},\ }\href {https://doi.org/10.1038/nphys1803} {\bibfield  {journal} {\bibinfo  {journal} {Nature Physics}\ }\textbf {\bibinfo {volume} {6}},\ \bibinfo {pages} {744} (\bibinfo {year} {2010})}\BibitemShut {NoStop}%
\bibitem [{\citenamefont {Shew}\ \emph {et~al.}(2011)\citenamefont {Shew}, \citenamefont {Yang}, \citenamefont {Yu}, \citenamefont {Roy},\ and\ \citenamefont {Plenz}}]{Shew2011}%
  \BibitemOpen
  \bibfield  {author} {\bibinfo {author} {\bibfnamefont {W.~L.}\ \bibnamefont {Shew}}, \bibinfo {author} {\bibfnamefont {H.}~\bibnamefont {Yang}}, \bibinfo {author} {\bibfnamefont {S.}~\bibnamefont {Yu}}, \bibinfo {author} {\bibfnamefont {R.}~\bibnamefont {Roy}},\ and\ \bibinfo {author} {\bibfnamefont {D.}~\bibnamefont {Plenz}},\ }\href {https://doi.org/10.1523/JNEUROSCI.4637-10.2011} {\bibfield  {journal} {\bibinfo  {journal} {The Journal of Neuroscience}\ }\textbf {\bibinfo {volume} {31}},\ \bibinfo {pages} {55} (\bibinfo {year} {2011})}\BibitemShut {NoStop}%
\bibitem [{\citenamefont {Shew}\ and\ \citenamefont {Plenz}(2013)}]{ShewPlenz2013}%
  \BibitemOpen
  \bibfield  {author} {\bibinfo {author} {\bibfnamefont {W.~L.}\ \bibnamefont {Shew}}\ and\ \bibinfo {author} {\bibfnamefont {D.}~\bibnamefont {Plenz}},\ }\href {https://doi.org/10.1177/1073858412445487} {\bibfield  {journal} {\bibinfo  {journal} {The Neuroscientist}\ }\textbf {\bibinfo {volume} {19}},\ \bibinfo {pages} {88} (\bibinfo {year} {2013})}\BibitemShut {NoStop}%
\bibitem [{\citenamefont {Petermann}\ \emph {et~al.}(2009)\citenamefont {Petermann}, \citenamefont {Thiagarajan}, \citenamefont {Lebedev}, \citenamefont {Nicolelis}, \citenamefont {Chialvo},\ and\ \citenamefont {Plenz}}]{Petermann2009}%
  \BibitemOpen
  \bibfield  {author} {\bibinfo {author} {\bibfnamefont {T.}~\bibnamefont {Petermann}}, \bibinfo {author} {\bibfnamefont {T.~C.}\ \bibnamefont {Thiagarajan}}, \bibinfo {author} {\bibfnamefont {M.~A.}\ \bibnamefont {Lebedev}}, \bibinfo {author} {\bibfnamefont {M.~A.~L.}\ \bibnamefont {Nicolelis}}, \bibinfo {author} {\bibfnamefont {D.~R.}\ \bibnamefont {Chialvo}},\ and\ \bibinfo {author} {\bibfnamefont {D.}~\bibnamefont {Plenz}},\ }\href {https://doi.org/10.1073/pnas.0904089106} {\bibfield  {journal} {\bibinfo  {journal} {Proceedings of the National Academy of Sciences}\ }\textbf {\bibinfo {volume} {106}},\ \bibinfo {pages} {15921} (\bibinfo {year} {2009})}\BibitemShut {NoStop}%
\bibitem [{\citenamefont {Shriki}\ \emph {et~al.}(2013)\citenamefont {Shriki}, \citenamefont {Alstott}, \citenamefont {Carver}, \citenamefont {Holroyd}, \citenamefont {Henson}, \citenamefont {Smith}, \citenamefont {Coppola}, \citenamefont {Bullmore},\ and\ \citenamefont {Plenz}}]{Shriki2013}%
  \BibitemOpen
  \bibfield  {author} {\bibinfo {author} {\bibfnamefont {O.}~\bibnamefont {Shriki}}, \bibinfo {author} {\bibfnamefont {J.}~\bibnamefont {Alstott}}, \bibinfo {author} {\bibfnamefont {F.}~\bibnamefont {Carver}}, \bibinfo {author} {\bibfnamefont {T.}~\bibnamefont {Holroyd}}, \bibinfo {author} {\bibfnamefont {R.~N.~A.}\ \bibnamefont {Henson}}, \bibinfo {author} {\bibfnamefont {M.~L.}\ \bibnamefont {Smith}}, \bibinfo {author} {\bibfnamefont {R.}~\bibnamefont {Coppola}}, \bibinfo {author} {\bibfnamefont {E.}~\bibnamefont {Bullmore}},\ and\ \bibinfo {author} {\bibfnamefont {D.}~\bibnamefont {Plenz}},\ }\href {https://doi.org/10.1523/JNEUROSCI.4286-12.2013} {\bibfield  {journal} {\bibinfo  {journal} {The Journal of Neuroscience}\ }\textbf {\bibinfo {volume} {33}},\ \bibinfo {pages} {7079} (\bibinfo {year} {2013})}\BibitemShut {NoStop}%
\bibitem [{\citenamefont {Legenstein}\ and\ \citenamefont {Maass}(2007)}]{Legenstein2007}%
  \BibitemOpen
  \bibfield  {author} {\bibinfo {author} {\bibfnamefont {R.}~\bibnamefont {Legenstein}}\ and\ \bibinfo {author} {\bibfnamefont {W.}~\bibnamefont {Maass}},\ }\href {https://doi.org/10.1016/j.neunet.2007.04.017} {\bibfield  {journal} {\bibinfo  {journal} {Neural Networks}\ }\textbf {\bibinfo {volume} {20}},\ \bibinfo {pages} {323} (\bibinfo {year} {2007})}\BibitemShut {NoStop}%
\bibitem [{\citenamefont {Mallinson}\ \emph {et~al.}(2019)\citenamefont {Mallinson}, \citenamefont {Shirai}, \citenamefont {Acharya}, \citenamefont {Bose}, \citenamefont {Galli},\ and\ \citenamefont {Brown}}]{Mallinson2019}%
  \BibitemOpen
  \bibfield  {author} {\bibinfo {author} {\bibfnamefont {J.~B.}\ \bibnamefont {Mallinson}}, \bibinfo {author} {\bibfnamefont {S.}~\bibnamefont {Shirai}}, \bibinfo {author} {\bibfnamefont {S.~K.}\ \bibnamefont {Acharya}}, \bibinfo {author} {\bibfnamefont {S.~K.}\ \bibnamefont {Bose}}, \bibinfo {author} {\bibfnamefont {E.}~\bibnamefont {Galli}},\ and\ \bibinfo {author} {\bibfnamefont {S.~A.}\ \bibnamefont {Brown}},\ }\href {https://doi.org/10.1126/sciadv.aaw8438} {\bibfield  {journal} {\bibinfo  {journal} {Science Advances}\ }\textbf {\bibinfo {volume} {5}},\ \bibinfo {pages} {eaaw8438} (\bibinfo {year} {2019})}\BibitemShut {NoStop}%
\bibitem [{\citenamefont {Zhu}\ \emph {et~al.}(2021)\citenamefont {Zhu}, \citenamefont {Hochstetter}, \citenamefont {Loeffler}, \citenamefont {Diaz-Alvarez}, \citenamefont {Nakayama}, \citenamefont {Lizier},\ and\ \citenamefont {Kuncic}}]{Zhu2021}%
  \BibitemOpen
  \bibfield  {author} {\bibinfo {author} {\bibfnamefont {R.}~\bibnamefont {Zhu}}, \bibinfo {author} {\bibfnamefont {J.}~\bibnamefont {Hochstetter}}, \bibinfo {author} {\bibfnamefont {A.}~\bibnamefont {Loeffler}}, \bibinfo {author} {\bibfnamefont {A.}~\bibnamefont {Diaz-Alvarez}}, \bibinfo {author} {\bibfnamefont {T.}~\bibnamefont {Nakayama}}, \bibinfo {author} {\bibfnamefont {J.~T.}\ \bibnamefont {Lizier}},\ and\ \bibinfo {author} {\bibfnamefont {Z.}~\bibnamefont {Kuncic}},\ }\href {https://doi.org/10.1038/s41598-021-92170-7} {\bibfield  {journal} {\bibinfo  {journal} {Scientific Reports}\ }\textbf {\bibinfo {volume} {11}},\ \bibinfo {pages} {13047} (\bibinfo {year} {2021})}\BibitemShut {NoStop}%
\bibitem [{\citenamefont {Jaeger}\ and\ \citenamefont {Haas}(2004)}]{Jaeger2004}%
  \BibitemOpen
  \bibfield  {author} {\bibinfo {author} {\bibfnamefont {H.}~\bibnamefont {Jaeger}}\ and\ \bibinfo {author} {\bibfnamefont {H.}~\bibnamefont {Haas}},\ }\href {https://doi.org/10.1126/science.1091277} {\bibfield  {journal} {\bibinfo  {journal} {Science}\ }\textbf {\bibinfo {volume} {304}},\ \bibinfo {pages} {78} (\bibinfo {year} {2004})}\BibitemShut {NoStop}%
\bibitem [{\citenamefont {Maass}\ \emph {et~al.}(2002)\citenamefont {Maass}, \citenamefont {Natschl{\"a}ger},\ and\ \citenamefont {Markram}}]{Maass2002}%
  \BibitemOpen
  \bibfield  {author} {\bibinfo {author} {\bibfnamefont {W.}~\bibnamefont {Maass}}, \bibinfo {author} {\bibfnamefont {T.}~\bibnamefont {Natschl{\"a}ger}},\ and\ \bibinfo {author} {\bibfnamefont {H.}~\bibnamefont {Markram}},\ }\href {https://doi.org/10.1162/089976602760407955} {\bibfield  {journal} {\bibinfo  {journal} {Neural Computation}\ }\textbf {\bibinfo {volume} {14}},\ \bibinfo {pages} {2531} (\bibinfo {year} {2002})}\BibitemShut {NoStop}%
\bibitem [{\citenamefont {Dambre}\ \emph {et~al.}(2012)\citenamefont {Dambre}, \citenamefont {Verstraeten}, \citenamefont {Schrauwen},\ and\ \citenamefont {Massar}}]{Dambre2012}%
  \BibitemOpen
  \bibfield  {author} {\bibinfo {author} {\bibfnamefont {J.}~\bibnamefont {Dambre}}, \bibinfo {author} {\bibfnamefont {D.}~\bibnamefont {Verstraeten}}, \bibinfo {author} {\bibfnamefont {B.}~\bibnamefont {Schrauwen}},\ and\ \bibinfo {author} {\bibfnamefont {S.}~\bibnamefont {Massar}},\ }\href {https://doi.org/10.1038/srep00514} {\bibfield  {journal} {\bibinfo  {journal} {Scientific Reports}\ }\textbf {\bibinfo {volume} {2}},\ \bibinfo {pages} {514} (\bibinfo {year} {2012})}\BibitemShut {NoStop}%
\bibitem [{\citenamefont {Boedecker}\ \emph {et~al.}(2012)\citenamefont {Boedecker}, \citenamefont {Obst}, \citenamefont {Lizier}, \citenamefont {Mayer},\ and\ \citenamefont {Asada}}]{Boedecker2012}%
  \BibitemOpen
  \bibfield  {author} {\bibinfo {author} {\bibfnamefont {J.}~\bibnamefont {Boedecker}}, \bibinfo {author} {\bibfnamefont {O.}~\bibnamefont {Obst}}, \bibinfo {author} {\bibfnamefont {J.~T.}\ \bibnamefont {Lizier}}, \bibinfo {author} {\bibfnamefont {N.~M.}\ \bibnamefont {Mayer}},\ and\ \bibinfo {author} {\bibfnamefont {M.}~\bibnamefont {Asada}},\ }\href {https://doi.org/10.1007/s12064-011-0146-8} {\bibfield  {journal} {\bibinfo  {journal} {Theory in Biosciences}\ }\textbf {\bibinfo {volume} {131}},\ \bibinfo {pages} {205} (\bibinfo {year} {2012})}\BibitemShut {NoStop}%
\bibitem [{\citenamefont {Grigoryeva}\ and\ \citenamefont {Ortega}(2018)}]{GrigoryevaOrtega2018}%
  \BibitemOpen
  \bibfield  {author} {\bibinfo {author} {\bibfnamefont {L.}~\bibnamefont {Grigoryeva}}\ and\ \bibinfo {author} {\bibfnamefont {J.-P.}\ \bibnamefont {Ortega}},\ }\href {https://doi.org/10.1016/j.neunet.2018.08.025} {\bibfield  {journal} {\bibinfo  {journal} {Neural Networks}\ }\textbf {\bibinfo {volume} {108}},\ \bibinfo {pages} {495} (\bibinfo {year} {2018})}\BibitemShut {NoStop}%
\bibitem [{\citenamefont {Hayden}\ and\ \citenamefont {Preskill}(2007)}]{Hayden2007}%
  \BibitemOpen
  \bibfield  {author} {\bibinfo {author} {\bibfnamefont {P.}~\bibnamefont {Hayden}}\ and\ \bibinfo {author} {\bibfnamefont {J.}~\bibnamefont {Preskill}},\ }\href {https://doi.org/10.1088/1126-6708/2007/09/120} {\bibfield  {journal} {\bibinfo  {journal} {Journal of High Energy Physics}\ }\textbf {\bibinfo {volume} {2007}},\ \bibinfo {pages} {120} (\bibinfo {year} {2007})}\BibitemShut {NoStop}%
\bibitem [{\citenamefont {Sekino}\ and\ \citenamefont {Susskind}(2008)}]{Sekino2008}%
  \BibitemOpen
  \bibfield  {author} {\bibinfo {author} {\bibfnamefont {Y.}~\bibnamefont {Sekino}}\ and\ \bibinfo {author} {\bibfnamefont {L.}~\bibnamefont {Susskind}},\ }\href {https://doi.org/10.1088/1126-6708/2008/10/065} {\bibfield  {journal} {\bibinfo  {journal} {Journal of High Energy Physics}\ }\textbf {\bibinfo {volume} {2008}},\ \bibinfo {pages} {065} (\bibinfo {year} {2008})}\BibitemShut {NoStop}%
\bibitem [{\citenamefont {Hosur}\ \emph {et~al.}(2016)\citenamefont {Hosur}, \citenamefont {Qi}, \citenamefont {Roberts},\ and\ \citenamefont {Yoshida}}]{Hosur2016}%
  \BibitemOpen
  \bibfield  {author} {\bibinfo {author} {\bibfnamefont {P.}~\bibnamefont {Hosur}}, \bibinfo {author} {\bibfnamefont {X.-L.}\ \bibnamefont {Qi}}, \bibinfo {author} {\bibfnamefont {D.~A.}\ \bibnamefont {Roberts}},\ and\ \bibinfo {author} {\bibfnamefont {B.}~\bibnamefont {Yoshida}},\ }\href {https://doi.org/10.1007/JHEP02(2016)004} {\bibfield  {journal} {\bibinfo  {journal} {Journal of High Energy Physics}\ }\textbf {\bibinfo {volume} {2016}},\ \bibinfo {pages} {4} (\bibinfo {year} {2016})}\BibitemShut {NoStop}%
\bibitem [{\citenamefont {Maldacena}\ \emph {et~al.}(2016)\citenamefont {Maldacena}, \citenamefont {Shenker},\ and\ \citenamefont {Stanford}}]{MSS2016}%
  \BibitemOpen
  \bibfield  {author} {\bibinfo {author} {\bibfnamefont {J.}~\bibnamefont {Maldacena}}, \bibinfo {author} {\bibfnamefont {S.~H.}\ \bibnamefont {Shenker}},\ and\ \bibinfo {author} {\bibfnamefont {D.}~\bibnamefont {Stanford}},\ }\href {https://doi.org/10.1007/JHEP08(2016)106} {\bibfield  {journal} {\bibinfo  {journal} {Journal of High Energy Physics}\ }\textbf {\bibinfo {volume} {2016}},\ \bibinfo {pages} {106} (\bibinfo {year} {2016})}\BibitemShut {NoStop}%
\bibitem [{\citenamefont {Swingle}(2018)}]{Swingle2018}%
  \BibitemOpen
  \bibfield  {author} {\bibinfo {author} {\bibfnamefont {B.}~\bibnamefont {Swingle}},\ }\href {https://doi.org/10.1038/s41567-018-0295-5} {\bibfield  {journal} {\bibinfo  {journal} {Nature Physics}\ }\textbf {\bibinfo {volume} {14}},\ \bibinfo {pages} {988} (\bibinfo {year} {2018})}\BibitemShut {NoStop}%
\bibitem [{\citenamefont {Xu}\ and\ \citenamefont {Swingle}(2024)}]{XuSwingle2024}%
  \BibitemOpen
  \bibfield  {author} {\bibinfo {author} {\bibfnamefont {S.}~\bibnamefont {Xu}}\ and\ \bibinfo {author} {\bibfnamefont {B.}~\bibnamefont {Swingle}},\ }\href {https://doi.org/10.1103/PRXQuantum.5.010201} {\bibfield  {journal} {\bibinfo  {journal} {PRX Quantum}\ }\textbf {\bibinfo {volume} {5}},\ \bibinfo {pages} {010201} (\bibinfo {year} {2024})}\BibitemShut {NoStop}%
\bibitem [{\citenamefont {Roberts}\ and\ \citenamefont {Yoshida}(2017)}]{RobertsYoshida2017}%
  \BibitemOpen
  \bibfield  {author} {\bibinfo {author} {\bibfnamefont {D.~A.}\ \bibnamefont {Roberts}}\ and\ \bibinfo {author} {\bibfnamefont {B.}~\bibnamefont {Yoshida}},\ }\href {https://doi.org/10.1007/JHEP04(2017)121} {\bibfield  {journal} {\bibinfo  {journal} {Journal of High Energy Physics}\ }\textbf {\bibinfo {volume} {2017}},\ \bibinfo {pages} {121} (\bibinfo {year} {2017})}\BibitemShut {NoStop}%
\bibitem [{\citenamefont {Dowling}\ \emph {et~al.}(2023)\citenamefont {Dowling}, \citenamefont {Kos},\ and\ \citenamefont {Modi}}]{Dowling2023}%
  \BibitemOpen
  \bibfield  {author} {\bibinfo {author} {\bibfnamefont {N.}~\bibnamefont {Dowling}}, \bibinfo {author} {\bibfnamefont {P.}~\bibnamefont {Kos}},\ and\ \bibinfo {author} {\bibfnamefont {K.}~\bibnamefont {Modi}},\ }\href {https://doi.org/10.1103/PhysRevLett.131.180403} {\bibfield  {journal} {\bibinfo  {journal} {Physical Review Letters}\ }\textbf {\bibinfo {volume} {131}},\ \bibinfo {pages} {180403} (\bibinfo {year} {2023})}\BibitemShut {NoStop}%
\bibitem [{\citenamefont {{Google Quantum AI and Collaborators}}(2025)}]{GoogleEchos}%
  \BibitemOpen
  \bibfield  {author} {\bibinfo {author} {\bibnamefont {{Google Quantum AI and Collaborators}}},\ }\href {https://doi.org/10.1038/s41586-025-09526-6} {\bibfield  {journal} {\bibinfo  {journal} {Nature}\ }\textbf {\bibinfo {volume} {646}},\ \bibinfo {pages} {825–830} (\bibinfo {year} {2025})}\BibitemShut {NoStop}%
\bibitem [{\citenamefont {Nakajima}\ \emph {et~al.}(2019)\citenamefont {Nakajima}, \citenamefont {Fujii}, \citenamefont {Negoro}, \citenamefont {Mitarai},\ and\ \citenamefont {Kitagawa}}]{Nakajima2019}%
  \BibitemOpen
  \bibfield  {author} {\bibinfo {author} {\bibfnamefont {K.}~\bibnamefont {Nakajima}}, \bibinfo {author} {\bibfnamefont {K.}~\bibnamefont {Fujii}}, \bibinfo {author} {\bibfnamefont {M.}~\bibnamefont {Negoro}}, \bibinfo {author} {\bibfnamefont {K.}~\bibnamefont {Mitarai}},\ and\ \bibinfo {author} {\bibfnamefont {M.}~\bibnamefont {Kitagawa}},\ }\href {https://doi.org/10.1103/PhysRevApplied.11.034021} {\bibfield  {journal} {\bibinfo  {journal} {Physical Review Applied}\ }\textbf {\bibinfo {volume} {11}},\ \bibinfo {pages} {034021} (\bibinfo {year} {2019})}\BibitemShut {NoStop}%
\bibitem [{\citenamefont {Fujii}\ and\ \citenamefont {Nakajima}(2017)}]{Fujii2017}%
  \BibitemOpen
  \bibfield  {author} {\bibinfo {author} {\bibfnamefont {K.}~\bibnamefont {Fujii}}\ and\ \bibinfo {author} {\bibfnamefont {K.}~\bibnamefont {Nakajima}},\ }\href {https://doi.org/10.1103/PhysRevApplied.8.024030} {\bibfield  {journal} {\bibinfo  {journal} {Physical Review Applied}\ }\textbf {\bibinfo {volume} {8}},\ \bibinfo {pages} {024030} (\bibinfo {year} {2017})}\BibitemShut {NoStop}%
\bibitem [{\citenamefont {Mujal}\ \emph {et~al.}(2021)\citenamefont {Mujal}, \citenamefont {Mart{\'i}nez-Pe{\~n}a}, \citenamefont {Nokkala}, \citenamefont {Garc{\'i}a-Beni}, \citenamefont {Giorgi}, \citenamefont {Soriano},\ and\ \citenamefont {Zambrini}}]{Mujal2021}%
  \BibitemOpen
  \bibfield  {author} {\bibinfo {author} {\bibfnamefont {P.}~\bibnamefont {Mujal}}, \bibinfo {author} {\bibfnamefont {R.}~\bibnamefont {Mart{\'i}nez-Pe{\~n}a}}, \bibinfo {author} {\bibfnamefont {J.}~\bibnamefont {Nokkala}}, \bibinfo {author} {\bibfnamefont {J.}~\bibnamefont {Garc{\'i}a-Beni}}, \bibinfo {author} {\bibfnamefont {G.~L.}\ \bibnamefont {Giorgi}}, \bibinfo {author} {\bibfnamefont {M.~C.}\ \bibnamefont {Soriano}},\ and\ \bibinfo {author} {\bibfnamefont {R.}~\bibnamefont {Zambrini}},\ }\href {https://doi.org/10.1002/qute.202100027} {\bibfield  {journal} {\bibinfo  {journal} {Advanced Quantum Technologies}\ }\textbf {\bibinfo {volume} {4}},\ \bibinfo {pages} {2100027} (\bibinfo {year} {2021})}\BibitemShut {NoStop}%
\bibitem [{\citenamefont {Kobayashi}\ \emph {et~al.}(2024)\citenamefont {Kobayashi}, \citenamefont {Tran},\ and\ \citenamefont {Nakajima}}]{Kobayashi2024}%
  \BibitemOpen
  \bibfield  {author} {\bibinfo {author} {\bibfnamefont {S.}~\bibnamefont {Kobayashi}}, \bibinfo {author} {\bibfnamefont {Q.~H.}\ \bibnamefont {Tran}},\ and\ \bibinfo {author} {\bibfnamefont {K.}~\bibnamefont {Nakajima}},\ }\href {https://doi.org/10.1103/PhysRevE.110.024207} {\bibfield  {journal} {\bibinfo  {journal} {Phys. Rev. E}\ }\textbf {\bibinfo {volume} {110}},\ \bibinfo {pages} {024207} (\bibinfo {year} {2024})}\BibitemShut {NoStop}%
\bibitem [{\citenamefont {Sannia}\ \emph {et~al.}(2024)\citenamefont {Sannia}, \citenamefont {Mart{\'i}nez-Pe{\~n}a}, \citenamefont {Soriano}, \citenamefont {Giorgi},\ and\ \citenamefont {Zambrini}}]{Sannia2024}%
  \BibitemOpen
  \bibfield  {author} {\bibinfo {author} {\bibfnamefont {A.}~\bibnamefont {Sannia}}, \bibinfo {author} {\bibfnamefont {R.}~\bibnamefont {Mart{\'i}nez-Pe{\~n}a}}, \bibinfo {author} {\bibfnamefont {M.~C.}\ \bibnamefont {Soriano}}, \bibinfo {author} {\bibfnamefont {G.~L.}\ \bibnamefont {Giorgi}},\ and\ \bibinfo {author} {\bibfnamefont {R.}~\bibnamefont {Zambrini}},\ }\href {https://doi.org/10.22331/q-2024-03-20-1291} {\bibfield  {journal} {\bibinfo  {journal} {Quantum}\ }\textbf {\bibinfo {volume} {8}},\ \bibinfo {pages} {1291} (\bibinfo {year} {2024})}\BibitemShut {NoStop}%
\bibitem [{\citenamefont {Hu}\ \emph {et~al.}(2024)\citenamefont {Hu}, \citenamefont {Khan}, \citenamefont {Bronn}, \citenamefont {Angelatos}, \citenamefont {Rowlands}, \citenamefont {Ribeill},\ and\ \citenamefont {T{\"u}reci}}]{Hu2024}%
  \BibitemOpen
  \bibfield  {author} {\bibinfo {author} {\bibfnamefont {F.}~\bibnamefont {Hu}}, \bibinfo {author} {\bibfnamefont {S.~A.}\ \bibnamefont {Khan}}, \bibinfo {author} {\bibfnamefont {N.~T.}\ \bibnamefont {Bronn}}, \bibinfo {author} {\bibfnamefont {G.}~\bibnamefont {Angelatos}}, \bibinfo {author} {\bibfnamefont {G.~E.}\ \bibnamefont {Rowlands}}, \bibinfo {author} {\bibfnamefont {G.~J.}\ \bibnamefont {Ribeill}},\ and\ \bibinfo {author} {\bibfnamefont {H.~E.}\ \bibnamefont {T{\"u}reci}},\ }\href {https://doi.org/10.1038/s41467-024-51162-7} {\bibfield  {journal} {\bibinfo  {journal} {Nature Communications}\ }\textbf {\bibinfo {volume} {15}},\ \bibinfo {pages} {7491} (\bibinfo {year} {2024})}\BibitemShut {NoStop}%
\bibitem [{\citenamefont {Xiong}\ \emph {et~al.}(2025)\citenamefont {Xiong}, \citenamefont {Holmes}, \citenamefont {Angrisani}, \citenamefont {Suzuki}, \citenamefont {Chotibut},\ and\ \citenamefont {Thanasilp}}]{Xiong2025}%
  \BibitemOpen
  \bibfield  {author} {\bibinfo {author} {\bibfnamefont {W.}~\bibnamefont {Xiong}}, \bibinfo {author} {\bibfnamefont {Z.}~\bibnamefont {Holmes}}, \bibinfo {author} {\bibfnamefont {A.}~\bibnamefont {Angrisani}}, \bibinfo {author} {\bibfnamefont {Y.}~\bibnamefont {Suzuki}}, \bibinfo {author} {\bibfnamefont {T.}~\bibnamefont {Chotibut}},\ and\ \bibinfo {author} {\bibfnamefont {S.}~\bibnamefont {Thanasilp}},\ }\href {https://doi.org/10.48550/arXiv.2505.10080} {\bibfield  {journal} {\bibinfo  {journal} {arXiv:2505.10080}\ ,\ \bibinfo {pages} {~}} (\bibinfo {year} {2025})}\BibitemShut {NoStop}%
\bibitem [{\citenamefont {Cenedese}\ \emph {et~al.}(2026)\citenamefont {Cenedese}, \citenamefont {Manzano}, \citenamefont {Giorgi},\ and\ \citenamefont {Zambrini}}]{Cenedese2026}%
  \BibitemOpen
  \bibfield  {author} {\bibinfo {author} {\bibfnamefont {G.}~\bibnamefont {Cenedese}}, \bibinfo {author} {\bibfnamefont {G.}~\bibnamefont {Manzano}}, \bibinfo {author} {\bibfnamefont {G.~L.}\ \bibnamefont {Giorgi}},\ and\ \bibinfo {author} {\bibfnamefont {R.}~\bibnamefont {Zambrini}},\ }\href {https://doi.org/10.48550/arXiv.2608.08279} {\bibfield  {journal} {\bibinfo  {journal} {arXiv:2608.08279}\ ,\ \bibinfo {pages} {~}} (\bibinfo {year} {2026})}\BibitemShut {NoStop}%
\bibitem [{\citenamefont {Negoro}\ \emph {et~al.}(2018)\citenamefont {Negoro}, \citenamefont {Mitarai}, \citenamefont {Fujii}, \citenamefont {Nakajima},\ and\ \citenamefont {Kitagawa}}]{Negoro2018}%
  \BibitemOpen
  \bibfield  {author} {\bibinfo {author} {\bibfnamefont {M.}~\bibnamefont {Negoro}}, \bibinfo {author} {\bibfnamefont {K.}~\bibnamefont {Mitarai}}, \bibinfo {author} {\bibfnamefont {K.}~\bibnamefont {Fujii}}, \bibinfo {author} {\bibfnamefont {K.}~\bibnamefont {Nakajima}},\ and\ \bibinfo {author} {\bibfnamefont {M.}~\bibnamefont {Kitagawa}},\ }\href {https://doi.org/10.48550/arXiv.1806.10910} {\bibfield  {journal} {\bibinfo  {journal} {arXiv:1806.10910}\ ,\ \bibinfo {pages} {~}} (\bibinfo {year} {2018})}\BibitemShut {NoStop}%
\bibitem [{\citenamefont {Hou}\ \emph {et~al.}(2026)\citenamefont {Hou}, \citenamefont {Hua}, \citenamefont {Wu}, \citenamefont {Xia}, \citenamefont {Chen}, \citenamefont {Li}, \citenamefont {Li}, \citenamefont {Peng},\ and\ \citenamefont {Du}}]{Hou2026}%
  \BibitemOpen
  \bibfield  {author} {\bibinfo {author} {\bibfnamefont {Y.}~\bibnamefont {Hou}}, \bibinfo {author} {\bibfnamefont {J.}~\bibnamefont {Hua}}, \bibinfo {author} {\bibfnamefont {Z.}~\bibnamefont {Wu}}, \bibinfo {author} {\bibfnamefont {W.}~\bibnamefont {Xia}}, \bibinfo {author} {\bibfnamefont {Y.}~\bibnamefont {Chen}}, \bibinfo {author} {\bibfnamefont {X.}~\bibnamefont {Li}}, \bibinfo {author} {\bibfnamefont {Z.}~\bibnamefont {Li}}, \bibinfo {author} {\bibfnamefont {X.}~\bibnamefont {Peng}},\ and\ \bibinfo {author} {\bibfnamefont {J.}~\bibnamefont {Du}},\ }\href {https://doi.org/10.1103/r8ww-qw7j} {\bibfield  {journal} {\bibinfo  {journal} {Physical Review Letters}\ }\textbf {\bibinfo {volume} {136}},\ \bibinfo {pages} {120602} (\bibinfo {year} {2026})}\BibitemShut {NoStop}%
\bibitem [{\citenamefont {Yasuda}\ \emph {et~al.}(2023)\citenamefont {Yasuda}, \citenamefont {Suzuki}, \citenamefont {Kubota}, \citenamefont {Nakajima}, \citenamefont {Gao}, \citenamefont {Zhang}, \citenamefont {Shimono}, \citenamefont {Nurdin},\ and\ \citenamefont {Yamamoto}}]{Yasuda2023}%
  \BibitemOpen
  \bibfield  {author} {\bibinfo {author} {\bibfnamefont {T.}~\bibnamefont {Yasuda}}, \bibinfo {author} {\bibfnamefont {Y.}~\bibnamefont {Suzuki}}, \bibinfo {author} {\bibfnamefont {T.}~\bibnamefont {Kubota}}, \bibinfo {author} {\bibfnamefont {K.}~\bibnamefont {Nakajima}}, \bibinfo {author} {\bibfnamefont {Q.}~\bibnamefont {Gao}}, \bibinfo {author} {\bibfnamefont {W.}~\bibnamefont {Zhang}}, \bibinfo {author} {\bibfnamefont {S.}~\bibnamefont {Shimono}}, \bibinfo {author} {\bibfnamefont {H.~I.}\ \bibnamefont {Nurdin}},\ and\ \bibinfo {author} {\bibfnamefont {N.}~\bibnamefont {Yamamoto}},\ }\href {https://doi.org/10.48550/arXiv.2310.06706} {\bibfield  {journal} {\bibinfo  {journal} {arXiv:2310.06706}\ ,\ \bibinfo {pages} {~}} (\bibinfo {year} {2023})}\BibitemShut {NoStop}%
\bibitem [{\citenamefont {Araiza~Bravo}\ \emph {et~al.}(2022)\citenamefont {Araiza~Bravo}, \citenamefont {Najafi}, \citenamefont {Gao},\ and\ \citenamefont {Yelin}}]{Bravo2022}%
  \BibitemOpen
  \bibfield  {author} {\bibinfo {author} {\bibfnamefont {R.}~\bibnamefont {Araiza~Bravo}}, \bibinfo {author} {\bibfnamefont {K.}~\bibnamefont {Najafi}}, \bibinfo {author} {\bibfnamefont {X.}~\bibnamefont {Gao}},\ and\ \bibinfo {author} {\bibfnamefont {S.~F.}\ \bibnamefont {Yelin}},\ }\href {https://doi.org/10.1103/PRXQuantum.3.030325} {\bibfield  {journal} {\bibinfo  {journal} {PRX Quantum}\ }\textbf {\bibinfo {volume} {3}},\ \bibinfo {pages} {030325} (\bibinfo {year} {2022})}\BibitemShut {NoStop}%
\bibitem [{\citenamefont {Kornja{\v{c}}a}\ \emph {et~al.}(2024)\citenamefont {Kornja{\v{c}}a}, \citenamefont {Hu}, \citenamefont {Zhao}, \citenamefont {Wurtz}, \citenamefont {Weinberg} \emph {et~al.}}]{Kornjaca2024}%
  \BibitemOpen
  \bibfield  {author} {\bibinfo {author} {\bibfnamefont {M.}~\bibnamefont {Kornja{\v{c}}a}}, \bibinfo {author} {\bibfnamefont {H.-Y.}\ \bibnamefont {Hu}}, \bibinfo {author} {\bibfnamefont {C.}~\bibnamefont {Zhao}}, \bibinfo {author} {\bibfnamefont {J.}~\bibnamefont {Wurtz}}, \bibinfo {author} {\bibfnamefont {P.}~\bibnamefont {Weinberg}}, \emph {et~al.},\ }\href {https://doi.org/10.48550/arXiv.2407.02553} {\bibfield  {journal} {\bibinfo  {journal} {arXiv:2407.02553}\ ,\ \bibinfo {pages} {~}} (\bibinfo {year} {2024})}\BibitemShut {NoStop}%
\bibitem [{\citenamefont {Mart{\'i}nez-Pe{\~n}a}\ \emph {et~al.}(2021)\citenamefont {Mart{\'i}nez-Pe{\~n}a}, \citenamefont {Giorgi}, \citenamefont {Nokkala}, \citenamefont {Soriano},\ and\ \citenamefont {Zambrini}}]{martinezpena2021}%
  \BibitemOpen
  \bibfield  {author} {\bibinfo {author} {\bibfnamefont {R.}~\bibnamefont {Mart{\'i}nez-Pe{\~n}a}}, \bibinfo {author} {\bibfnamefont {G.~L.}\ \bibnamefont {Giorgi}}, \bibinfo {author} {\bibfnamefont {J.}~\bibnamefont {Nokkala}}, \bibinfo {author} {\bibfnamefont {M.~C.}\ \bibnamefont {Soriano}},\ and\ \bibinfo {author} {\bibfnamefont {R.}~\bibnamefont {Zambrini}},\ }\href {https://doi.org/10.1103/PhysRevLett.127.100502} {\bibfield  {journal} {\bibinfo  {journal} {Physical Review Letters}\ }\textbf {\bibinfo {volume} {127}},\ \bibinfo {pages} {100502} (\bibinfo {year} {2021})}\BibitemShut {NoStop}%
\bibitem [{\citenamefont {Kobayashi}\ and\ \citenamefont {Motome}(2025)}]{Kobayashi2025probe}%
  \BibitemOpen
  \bibfield  {author} {\bibinfo {author} {\bibfnamefont {K.}~\bibnamefont {Kobayashi}}\ and\ \bibinfo {author} {\bibfnamefont {Y.}~\bibnamefont {Motome}},\ }\href {https://doi.org/10.1038/s41467-025-58751-0} {\bibfield  {journal} {\bibinfo  {journal} {Nature Communications}\ }\textbf {\bibinfo {volume} {16}},\ \bibinfo {pages} {3871} (\bibinfo {year} {2025})}\BibitemShut {NoStop}%
\bibitem [{\citenamefont {Kobayashi}\ and\ \citenamefont {Motome}(2026)}]{Kobayashi2026edge}%
  \BibitemOpen
  \bibfield  {author} {\bibinfo {author} {\bibfnamefont {K.}~\bibnamefont {Kobayashi}}\ and\ \bibinfo {author} {\bibfnamefont {Y.}~\bibnamefont {Motome}},\ }\href {https://doi.org/10.1103/j2qj-vwcl} {\bibfield  {journal} {\bibinfo  {journal} {Physical Review Letters}\ }\textbf {\bibinfo {volume} {136}},\ \bibinfo {pages} {040602} (\bibinfo {year} {2026})}\BibitemShut {NoStop}%
\bibitem [{\citenamefont {Bermejo}\ \emph {et~al.}(2026)\citenamefont {Bermejo}, \citenamefont {Villalonga}, \citenamefont {Ware}, \citenamefont {Vidal},\ and\ \citenamefont {Szasz}}]{bermejo2026tensor}%
  \BibitemOpen
  \bibfield  {author} {\bibinfo {author} {\bibfnamefont {P.}~\bibnamefont {Bermejo}}, \bibinfo {author} {\bibfnamefont {B.}~\bibnamefont {Villalonga}}, \bibinfo {author} {\bibfnamefont {B.}~\bibnamefont {Ware}}, \bibinfo {author} {\bibfnamefont {G.}~\bibnamefont {Vidal}},\ and\ \bibinfo {author} {\bibfnamefont {A.}~\bibnamefont {Szasz}},\ }\href {https://doi.org/10.48550/arXiv.2604.15427} {\bibfield  {journal} {\bibinfo  {journal} {arXiv:2604.15427}\ ,\ \bibinfo {pages} {~}} (\bibinfo {year} {2026})}\BibitemShut {NoStop}%
\bibitem [{\citenamefont {Hu}\ \emph {et~al.}(2023)\citenamefont {Hu}, \citenamefont {Angelatos}, \citenamefont {Khan}, \citenamefont {Vives}, \citenamefont {T{\"u}reci}, \citenamefont {Bello}, \citenamefont {Rowlands}, \citenamefont {Ribeill},\ and\ \citenamefont {T{\"u}reci}}]{Hu2023REC}%
  \BibitemOpen
  \bibfield  {author} {\bibinfo {author} {\bibfnamefont {F.}~\bibnamefont {Hu}}, \bibinfo {author} {\bibfnamefont {G.}~\bibnamefont {Angelatos}}, \bibinfo {author} {\bibfnamefont {S.~A.}\ \bibnamefont {Khan}}, \bibinfo {author} {\bibfnamefont {M.}~\bibnamefont {Vives}}, \bibinfo {author} {\bibfnamefont {E.}~\bibnamefont {T{\"u}reci}}, \bibinfo {author} {\bibfnamefont {L.}~\bibnamefont {Bello}}, \bibinfo {author} {\bibfnamefont {G.~E.}\ \bibnamefont {Rowlands}}, \bibinfo {author} {\bibfnamefont {G.~J.}\ \bibnamefont {Ribeill}},\ and\ \bibinfo {author} {\bibfnamefont {H.~E.}\ \bibnamefont {T{\"u}reci}},\ }\href {https://doi.org/10.1103/PhysRevX.13.041020} {\bibfield  {journal} {\bibinfo  {journal} {Physical Review X}\ }\textbf {\bibinfo {volume} {13}},\ \bibinfo {pages} {041020} (\bibinfo {year} {2023})}\BibitemShut {NoStop}%
\bibitem [{\citenamefont {Schuld}\ \emph {et~al.}(2021)\citenamefont {Schuld}, \citenamefont {Sweke},\ and\ \citenamefont {Meyer}}]{Schuld2021}%
  \BibitemOpen
  \bibfield  {author} {\bibinfo {author} {\bibfnamefont {M.}~\bibnamefont {Schuld}}, \bibinfo {author} {\bibfnamefont {R.}~\bibnamefont {Sweke}},\ and\ \bibinfo {author} {\bibfnamefont {J.~J.}\ \bibnamefont {Meyer}},\ }\href {https://doi.org/10.1103/PhysRevA.103.032430} {\bibfield  {journal} {\bibinfo  {journal} {Physical Review A}\ }\textbf {\bibinfo {volume} {103}},\ \bibinfo {pages} {032430} (\bibinfo {year} {2021})}\BibitemShut {NoStop}%
\bibitem [{\citenamefont {Innocenti}\ \emph {et~al.}(2023)\citenamefont {Innocenti}, \citenamefont {Lorenzo}, \citenamefont {Palmisano}, \citenamefont {Ferraro}, \citenamefont {Paternostro},\ and\ \citenamefont {Palma}}]{Innocenti2023}%
  \BibitemOpen
  \bibfield  {author} {\bibinfo {author} {\bibfnamefont {L.}~\bibnamefont {Innocenti}}, \bibinfo {author} {\bibfnamefont {S.}~\bibnamefont {Lorenzo}}, \bibinfo {author} {\bibfnamefont {I.}~\bibnamefont {Palmisano}}, \bibinfo {author} {\bibfnamefont {A.}~\bibnamefont {Ferraro}}, \bibinfo {author} {\bibfnamefont {M.}~\bibnamefont {Paternostro}},\ and\ \bibinfo {author} {\bibfnamefont {G.~M.}\ \bibnamefont {Palma}},\ }\href {https://doi.org/10.1038/s42005-023-01233-w} {\bibfield  {journal} {\bibinfo  {journal} {Communications Physics}\ }\textbf {\bibinfo {volume} {6}},\ \bibinfo {pages} {118} (\bibinfo {year} {2023})}\BibitemShut {NoStop}%
\bibitem [{\citenamefont {Sch{\"u}tte}\ \emph {et~al.}(2025)\citenamefont {Sch{\"u}tte}, \citenamefont {G{\"o}tting}, \citenamefont {M{\"u}ntinga}, \citenamefont {Gies},\ and\ \citenamefont {List}}]{Schutte2025}%
  \BibitemOpen
  \bibfield  {author} {\bibinfo {author} {\bibfnamefont {N.-E.}\ \bibnamefont {Sch{\"u}tte}}, \bibinfo {author} {\bibfnamefont {N.}~\bibnamefont {G{\"o}tting}}, \bibinfo {author} {\bibfnamefont {H.}~\bibnamefont {M{\"u}ntinga}}, \bibinfo {author} {\bibfnamefont {C.}~\bibnamefont {Gies}},\ and\ \bibinfo {author} {\bibfnamefont {M.}~\bibnamefont {List}},\ }\href {https://doi.org/10.48550/arXiv.2501.15528} {\bibfield  {journal} {\bibinfo  {journal} {arXiv:2501.15528}\ ,\ \bibinfo {pages} {~}} (\bibinfo {year} {2025})}\BibitemShut {NoStop}%
\bibitem [{\citenamefont {Ebadi}\ \emph {et~al.}(2021)\citenamefont {Ebadi}, \citenamefont {Wang}, \citenamefont {Levine}, \citenamefont {Keesling}, \citenamefont {Semeghini}, \citenamefont {Omran}, \citenamefont {Bluvstein}, \citenamefont {Samajdar}, \citenamefont {Pichler}, \citenamefont {Ho}, \citenamefont {Choi}, \citenamefont {Sachdev}, \citenamefont {Greiner}, \citenamefont {Vuleti{\'c}},\ and\ \citenamefont {Lukin}}]{Ebadi2021}%
  \BibitemOpen
  \bibfield  {author} {\bibinfo {author} {\bibfnamefont {S.}~\bibnamefont {Ebadi}}, \bibinfo {author} {\bibfnamefont {T.~T.}\ \bibnamefont {Wang}}, \bibinfo {author} {\bibfnamefont {H.}~\bibnamefont {Levine}}, \bibinfo {author} {\bibfnamefont {A.}~\bibnamefont {Keesling}}, \bibinfo {author} {\bibfnamefont {G.}~\bibnamefont {Semeghini}}, \bibinfo {author} {\bibfnamefont {A.}~\bibnamefont {Omran}}, \bibinfo {author} {\bibfnamefont {D.}~\bibnamefont {Bluvstein}}, \bibinfo {author} {\bibfnamefont {R.}~\bibnamefont {Samajdar}}, \bibinfo {author} {\bibfnamefont {H.}~\bibnamefont {Pichler}}, \bibinfo {author} {\bibfnamefont {W.~W.}\ \bibnamefont {Ho}}, \bibinfo {author} {\bibfnamefont {S.}~\bibnamefont {Choi}}, \bibinfo {author} {\bibfnamefont {S.}~\bibnamefont {Sachdev}}, \bibinfo {author} {\bibfnamefont {M.}~\bibnamefont {Greiner}}, \bibinfo {author} {\bibfnamefont {V.}~\bibnamefont {Vuleti{\'c}}},\ and\ \bibinfo {author} {\bibfnamefont {M.~D.}\ \bibnamefont {Lukin}},\ }\href
  {https://doi.org/10.1038/s41586-021-03582-4} {\bibfield  {journal} {\bibinfo  {journal} {Nature}\ }\textbf {\bibinfo {volume} {595}},\ \bibinfo {pages} {227} (\bibinfo {year} {2021})}\BibitemShut {NoStop}%
\bibitem [{\citenamefont {Scholl}\ \emph {et~al.}(2021)\citenamefont {Scholl}, \citenamefont {Schuler}, \citenamefont {Williams}, \citenamefont {Eberharter}, \citenamefont {Barredo}, \citenamefont {Schymik}, \citenamefont {Lienhard}, \citenamefont {Henry}, \citenamefont {Lang}, \citenamefont {Lahaye}, \citenamefont {L{\"a}uchli},\ and\ \citenamefont {Browaeys}}]{Scholl2021}%
  \BibitemOpen
  \bibfield  {author} {\bibinfo {author} {\bibfnamefont {P.}~\bibnamefont {Scholl}}, \bibinfo {author} {\bibfnamefont {M.}~\bibnamefont {Schuler}}, \bibinfo {author} {\bibfnamefont {H.~J.}\ \bibnamefont {Williams}}, \bibinfo {author} {\bibfnamefont {A.~A.}\ \bibnamefont {Eberharter}}, \bibinfo {author} {\bibfnamefont {D.}~\bibnamefont {Barredo}}, \bibinfo {author} {\bibfnamefont {K.-N.}\ \bibnamefont {Schymik}}, \bibinfo {author} {\bibfnamefont {V.}~\bibnamefont {Lienhard}}, \bibinfo {author} {\bibfnamefont {L.-P.}\ \bibnamefont {Henry}}, \bibinfo {author} {\bibfnamefont {T.~C.}\ \bibnamefont {Lang}}, \bibinfo {author} {\bibfnamefont {T.}~\bibnamefont {Lahaye}}, \bibinfo {author} {\bibfnamefont {A.~M.}\ \bibnamefont {L{\"a}uchli}},\ and\ \bibinfo {author} {\bibfnamefont {A.}~\bibnamefont {Browaeys}},\ }\href {https://doi.org/10.1038/s41586-021-03585-1} {\bibfield  {journal} {\bibinfo  {journal} {Nature}\ }\textbf {\bibinfo {volume} {595}},\ \bibinfo {pages} {233} (\bibinfo {year} {2021})}\BibitemShut
  {NoStop}%
\bibitem [{\citenamefont {Krantz}\ \emph {et~al.}(2019)\citenamefont {Krantz}, \citenamefont {Kjaergaard}, \citenamefont {Yan}, \citenamefont {Orlando}, \citenamefont {Gustavsson},\ and\ \citenamefont {Oliver}}]{Krantz2019}%
  \BibitemOpen
  \bibfield  {author} {\bibinfo {author} {\bibfnamefont {P.}~\bibnamefont {Krantz}}, \bibinfo {author} {\bibfnamefont {M.}~\bibnamefont {Kjaergaard}}, \bibinfo {author} {\bibfnamefont {F.}~\bibnamefont {Yan}}, \bibinfo {author} {\bibfnamefont {T.~P.}\ \bibnamefont {Orlando}}, \bibinfo {author} {\bibfnamefont {S.}~\bibnamefont {Gustavsson}},\ and\ \bibinfo {author} {\bibfnamefont {W.~D.}\ \bibnamefont {Oliver}},\ }\href {https://doi.org/10.1063/1.5089550} {\bibfield  {journal} {\bibinfo  {journal} {Applied Physics Reviews}\ }\textbf {\bibinfo {volume} {6}},\ \bibinfo {pages} {021318} (\bibinfo {year} {2019})}\BibitemShut {NoStop}%
\bibitem [{\citenamefont {Johnson}\ \emph {et~al.}(2011)\citenamefont {Johnson}, \citenamefont {Amin}, \citenamefont {Gildert}, \citenamefont {Lanting}, \citenamefont {Hamze}, \citenamefont {Dickson}, \citenamefont {Harris}, \citenamefont {Berkley}, \citenamefont {Johansson}, \citenamefont {Bunyk} \emph {et~al.}}]{Johnson2011}%
  \BibitemOpen
  \bibfield  {author} {\bibinfo {author} {\bibfnamefont {M.~W.}\ \bibnamefont {Johnson}}, \bibinfo {author} {\bibfnamefont {M.~H.~S.}\ \bibnamefont {Amin}}, \bibinfo {author} {\bibfnamefont {S.}~\bibnamefont {Gildert}}, \bibinfo {author} {\bibfnamefont {T.}~\bibnamefont {Lanting}}, \bibinfo {author} {\bibfnamefont {F.}~\bibnamefont {Hamze}}, \bibinfo {author} {\bibfnamefont {N.}~\bibnamefont {Dickson}}, \bibinfo {author} {\bibfnamefont {R.}~\bibnamefont {Harris}}, \bibinfo {author} {\bibfnamefont {A.~J.}\ \bibnamefont {Berkley}}, \bibinfo {author} {\bibfnamefont {J.}~\bibnamefont {Johansson}}, \bibinfo {author} {\bibfnamefont {P.}~\bibnamefont {Bunyk}}, \emph {et~al.},\ }\href {https://doi.org/10.1038/nature10012} {\bibfield  {journal} {\bibinfo  {journal} {Nature}\ }\textbf {\bibinfo {volume} {473}},\ \bibinfo {pages} {194} (\bibinfo {year} {2011})}\BibitemShut {NoStop}%
\bibitem [{\citenamefont {Settino}\ \emph {et~al.}(2025)\citenamefont {Settino}, \citenamefont {Salatino}, \citenamefont {Mariani}, \citenamefont {D'Amore}, \citenamefont {Channab}, \citenamefont {Bozzolo}, \citenamefont {Vallisa}, \citenamefont {Barill{\`a}}, \citenamefont {Policicchio}, \citenamefont {Lo~Gullo}, \citenamefont {Giordano}, \citenamefont {Mastroianni},\ and\ \citenamefont {Plastina}}]{Settino2025}%
  \BibitemOpen
  \bibfield  {author} {\bibinfo {author} {\bibfnamefont {J.}~\bibnamefont {Settino}}, \bibinfo {author} {\bibfnamefont {L.}~\bibnamefont {Salatino}}, \bibinfo {author} {\bibfnamefont {L.}~\bibnamefont {Mariani}}, \bibinfo {author} {\bibfnamefont {F.}~\bibnamefont {D'Amore}}, \bibinfo {author} {\bibfnamefont {M.}~\bibnamefont {Channab}}, \bibinfo {author} {\bibfnamefont {L.}~\bibnamefont {Bozzolo}}, \bibinfo {author} {\bibfnamefont {S.}~\bibnamefont {Vallisa}}, \bibinfo {author} {\bibfnamefont {P.}~\bibnamefont {Barill{\`a}}}, \bibinfo {author} {\bibfnamefont {A.}~\bibnamefont {Policicchio}}, \bibinfo {author} {\bibfnamefont {N.}~\bibnamefont {Lo~Gullo}}, \bibinfo {author} {\bibfnamefont {A.}~\bibnamefont {Giordano}}, \bibinfo {author} {\bibfnamefont {C.}~\bibnamefont {Mastroianni}},\ and\ \bibinfo {author} {\bibfnamefont {F.}~\bibnamefont {Plastina}},\ }\href {https://doi.org/10.1103/wzwv-7rk2} {\bibfield  {journal} {\bibinfo  {journal} {Physical Review Applied}\ }\textbf {\bibinfo {volume} {24}},\ \bibinfo
  {pages} {024019} (\bibinfo {year} {2025})}\BibitemShut {NoStop}%
\bibitem [{\citenamefont {Keenan}\ and\ \citenamefont {Zambrini}(2026)}]{Keenan2026}%
  \BibitemOpen
  \bibfield  {author} {\bibinfo {author} {\bibfnamefont {N.}~\bibnamefont {Keenan}}\ and\ \bibinfo {author} {\bibfnamefont {R.}~\bibnamefont {Zambrini}},\ }\href {https://doi.org/10.48550/arXiv.2608.07677} {\bibfield  {journal} {\bibinfo  {journal} {arXiv:2608.07677}\ ,\ \bibinfo {pages} {~}} (\bibinfo {year} {2026})}\BibitemShut {NoStop}%
\bibitem [{\citenamefont {Fan}\ \emph {et~al.}(2017)\citenamefont {Fan}, \citenamefont {Zhang}, \citenamefont {Shen},\ and\ \citenamefont {Zhai}}]{Fan2017}%
  \BibitemOpen
  \bibfield  {author} {\bibinfo {author} {\bibfnamefont {R.}~\bibnamefont {Fan}}, \bibinfo {author} {\bibfnamefont {P.}~\bibnamefont {Zhang}}, \bibinfo {author} {\bibfnamefont {H.}~\bibnamefont {Shen}},\ and\ \bibinfo {author} {\bibfnamefont {H.}~\bibnamefont {Zhai}},\ }\href {https://doi.org/10.1016/j.scib.2017.04.011} {\bibfield  {journal} {\bibinfo  {journal} {Science Bulletin}\ }\textbf {\bibinfo {volume} {62}},\ \bibinfo {pages} {707} (\bibinfo {year} {2017})}\BibitemShut {NoStop}%
\bibitem [{\citenamefont {Ba{\~n}uls}\ \emph {et~al.}(2011)\citenamefont {Ba{\~n}uls}, \citenamefont {Cirac},\ and\ \citenamefont {Hastings}}]{Banuls2011}%
  \BibitemOpen
  \bibfield  {author} {\bibinfo {author} {\bibfnamefont {M.~C.}\ \bibnamefont {Ba{\~n}uls}}, \bibinfo {author} {\bibfnamefont {J.~I.}\ \bibnamefont {Cirac}},\ and\ \bibinfo {author} {\bibfnamefont {M.~B.}\ \bibnamefont {Hastings}},\ }\href {https://doi.org/10.1103/PhysRevLett.106.050405} {\bibfield  {journal} {\bibinfo  {journal} {Physical Review Letters}\ }\textbf {\bibinfo {volume} {106}},\ \bibinfo {pages} {050405} (\bibinfo {year} {2011})}\BibitemShut {NoStop}%
\bibitem [{\citenamefont {Kim}\ and\ \citenamefont {Huse}(2013)}]{KimHuse2013}%
  \BibitemOpen
  \bibfield  {author} {\bibinfo {author} {\bibfnamefont {H.}~\bibnamefont {Kim}}\ and\ \bibinfo {author} {\bibfnamefont {D.~A.}\ \bibnamefont {Huse}},\ }\href {https://doi.org/10.1103/PhysRevLett.111.127205} {\bibfield  {journal} {\bibinfo  {journal} {Physical Review Letters}\ }\textbf {\bibinfo {volume} {111}},\ \bibinfo {pages} {127205} (\bibinfo {year} {2013})}\BibitemShut {NoStop}%
\bibitem [{\citenamefont {Tsukerman}(2017)}]{Tsukerman2017}%
  \BibitemOpen
  \bibfield  {author} {\bibinfo {author} {\bibfnamefont {E.}~\bibnamefont {Tsukerman}},\ }\href {https://doi.org/10.1103/PhysRevB.95.115121} {\bibfield  {journal} {\bibinfo  {journal} {Phys. Rev. B}\ }\textbf {\bibinfo {volume} {95}},\ \bibinfo {pages} {115121} (\bibinfo {year} {2017})}\BibitemShut {NoStop}%
\bibitem [{\citenamefont {Mackey}\ and\ \citenamefont {Glass}(1977)}]{mackey1977oscillation}%
  \BibitemOpen
  \bibfield  {author} {\bibinfo {author} {\bibfnamefont {M.~C.}\ \bibnamefont {Mackey}}\ and\ \bibinfo {author} {\bibfnamefont {L.}~\bibnamefont {Glass}},\ }\href {https://doi.org/10.1126/science.267326} {\bibfield  {journal} {\bibinfo  {journal} {Science}\ }\textbf {\bibinfo {volume} {197}},\ \bibinfo {pages} {287} (\bibinfo {year} {1977})}\BibitemShut {NoStop}%
\bibitem [{\citenamefont {Chen}\ and\ \citenamefont {Nurdin}(2019)}]{ChenNurdin2019}%
  \BibitemOpen
  \bibfield  {author} {\bibinfo {author} {\bibfnamefont {J.}~\bibnamefont {Chen}}\ and\ \bibinfo {author} {\bibfnamefont {H.~I.}\ \bibnamefont {Nurdin}},\ }\href {https://doi.org/10.1007/s11128-019-2311-9} {\bibfield  {journal} {\bibinfo  {journal} {Quantum Information Processing}\ }\textbf {\bibinfo {volume} {18}},\ \bibinfo {pages} {198} (\bibinfo {year} {2019})}\BibitemShut {NoStop}%
\bibitem [{\citenamefont {Kubota}\ \emph {et~al.}(2023)\citenamefont {Kubota}, \citenamefont {Suzuki}, \citenamefont {Kobayashi}, \citenamefont {Tran}, \citenamefont {Yamamoto},\ and\ \citenamefont {Nakajima}}]{Kubota2023}%
  \BibitemOpen
  \bibfield  {author} {\bibinfo {author} {\bibfnamefont {T.}~\bibnamefont {Kubota}}, \bibinfo {author} {\bibfnamefont {Y.}~\bibnamefont {Suzuki}}, \bibinfo {author} {\bibfnamefont {S.}~\bibnamefont {Kobayashi}}, \bibinfo {author} {\bibfnamefont {Q.~H.}\ \bibnamefont {Tran}}, \bibinfo {author} {\bibfnamefont {N.}~\bibnamefont {Yamamoto}},\ and\ \bibinfo {author} {\bibfnamefont {K.}~\bibnamefont {Nakajima}},\ }\href {https://doi.org/10.1103/PhysRevResearch.5.023057} {\bibfield  {journal} {\bibinfo  {journal} {Physical Review Research}\ }\textbf {\bibinfo {volume} {5}},\ \bibinfo {pages} {023057} (\bibinfo {year} {2023})}\BibitemShut {NoStop}%
\bibitem [{\citenamefont {Kaplan}\ \emph {et~al.}(2020)\citenamefont {Kaplan}, \citenamefont {McCandlish}, \citenamefont {Henighan}, \citenamefont {Brown}, \citenamefont {Chess}, \citenamefont {Child}, \citenamefont {Gray}, \citenamefont {Radford}, \citenamefont {Wu},\ and\ \citenamefont {Amodei}}]{Kaplan2020}%
  \BibitemOpen
  \bibfield  {author} {\bibinfo {author} {\bibfnamefont {J.}~\bibnamefont {Kaplan}}, \bibinfo {author} {\bibfnamefont {S.}~\bibnamefont {McCandlish}}, \bibinfo {author} {\bibfnamefont {T.}~\bibnamefont {Henighan}}, \bibinfo {author} {\bibfnamefont {T.~B.}\ \bibnamefont {Brown}}, \bibinfo {author} {\bibfnamefont {B.}~\bibnamefont {Chess}}, \bibinfo {author} {\bibfnamefont {R.}~\bibnamefont {Child}}, \bibinfo {author} {\bibfnamefont {S.}~\bibnamefont {Gray}}, \bibinfo {author} {\bibfnamefont {A.}~\bibnamefont {Radford}}, \bibinfo {author} {\bibfnamefont {J.}~\bibnamefont {Wu}},\ and\ \bibinfo {author} {\bibfnamefont {D.}~\bibnamefont {Amodei}},\ }\href {https://doi.org/10.48550/arXiv.2001.08361} {\bibfield  {journal} {\bibinfo  {journal} {arXiv:2001.08361}\ ,\ \bibinfo {pages} {~}} (\bibinfo {year} {2020})}\BibitemShut {NoStop}%
\bibitem [{\citenamefont {Hoffmann}\ \emph {et~al.}(2022)\citenamefont {Hoffmann}, \citenamefont {Borgeaud}, \citenamefont {Mensch}, \citenamefont {Buchatskaya}, \citenamefont {Cai}, \citenamefont {Rutherford}, \citenamefont {de~Las~Casas}, \citenamefont {Hendricks}, \citenamefont {Welbl}, \citenamefont {Clark} \emph {et~al.}}]{Hoffmann2022}%
  \BibitemOpen
  \bibfield  {author} {\bibinfo {author} {\bibfnamefont {J.}~\bibnamefont {Hoffmann}}, \bibinfo {author} {\bibfnamefont {S.}~\bibnamefont {Borgeaud}}, \bibinfo {author} {\bibfnamefont {A.}~\bibnamefont {Mensch}}, \bibinfo {author} {\bibfnamefont {E.}~\bibnamefont {Buchatskaya}}, \bibinfo {author} {\bibfnamefont {T.}~\bibnamefont {Cai}}, \bibinfo {author} {\bibfnamefont {E.}~\bibnamefont {Rutherford}}, \bibinfo {author} {\bibfnamefont {D.}~\bibnamefont {de~Las~Casas}}, \bibinfo {author} {\bibfnamefont {L.~A.}\ \bibnamefont {Hendricks}}, \bibinfo {author} {\bibfnamefont {J.}~\bibnamefont {Welbl}}, \bibinfo {author} {\bibfnamefont {A.}~\bibnamefont {Clark}}, \emph {et~al.},\ }\href {https://doi.org/10.48550/arXiv.2203.15556} {\bibfield  {journal} {\bibinfo  {journal} {arXiv:2203.15556}\ ,\ \bibinfo {pages} {~}} (\bibinfo {year} {2022})}\BibitemShut {NoStop}%
\bibitem [{\citenamefont {Wright}\ \emph {et~al.}(2022)\citenamefont {Wright}, \citenamefont {Onodera}, \citenamefont {Stein}, \citenamefont {Wang}, \citenamefont {Schachter}, \citenamefont {Hu},\ and\ \citenamefont {McMahon}}]{Wright2022}%
  \BibitemOpen
  \bibfield  {author} {\bibinfo {author} {\bibfnamefont {L.~G.}\ \bibnamefont {Wright}}, \bibinfo {author} {\bibfnamefont {T.}~\bibnamefont {Onodera}}, \bibinfo {author} {\bibfnamefont {M.~M.}\ \bibnamefont {Stein}}, \bibinfo {author} {\bibfnamefont {T.}~\bibnamefont {Wang}}, \bibinfo {author} {\bibfnamefont {D.~T.}\ \bibnamefont {Schachter}}, \bibinfo {author} {\bibfnamefont {Z.}~\bibnamefont {Hu}},\ and\ \bibinfo {author} {\bibfnamefont {P.~L.}\ \bibnamefont {McMahon}},\ }\href {https://doi.org/10.1038/s41586-021-04223-6} {\bibfield  {journal} {\bibinfo  {journal} {Nature}\ }\textbf {\bibinfo {volume} {601}},\ \bibinfo {pages} {549} (\bibinfo {year} {2022})}\BibitemShut {NoStop}%
\bibitem [{\citenamefont {Momeni}\ \emph {et~al.}(2023)\citenamefont {Momeni}, \citenamefont {Rahmani}, \citenamefont {Mall{\'e}jac}, \citenamefont {del Hougne},\ and\ \citenamefont {Fleury}}]{Momeni2023}%
  \BibitemOpen
  \bibfield  {author} {\bibinfo {author} {\bibfnamefont {A.}~\bibnamefont {Momeni}}, \bibinfo {author} {\bibfnamefont {B.}~\bibnamefont {Rahmani}}, \bibinfo {author} {\bibfnamefont {M.}~\bibnamefont {Mall{\'e}jac}}, \bibinfo {author} {\bibfnamefont {P.}~\bibnamefont {del Hougne}},\ and\ \bibinfo {author} {\bibfnamefont {R.}~\bibnamefont {Fleury}},\ }\href {https://doi.org/10.1126/science.adi8474} {\bibfield  {journal} {\bibinfo  {journal} {Science}\ }\textbf {\bibinfo {volume} {382}},\ \bibinfo {pages} {1297} (\bibinfo {year} {2023})}\BibitemShut {NoStop}%
\bibitem [{\citenamefont {Markovi{\'c}}\ \emph {et~al.}(2020)\citenamefont {Markovi{\'c}}, \citenamefont {Mizrahi}, \citenamefont {Querlioz},\ and\ \citenamefont {Grollier}}]{Markovic2020}%
  \BibitemOpen
  \bibfield  {author} {\bibinfo {author} {\bibfnamefont {D.}~\bibnamefont {Markovi{\'c}}}, \bibinfo {author} {\bibfnamefont {A.}~\bibnamefont {Mizrahi}}, \bibinfo {author} {\bibfnamefont {D.}~\bibnamefont {Querlioz}},\ and\ \bibinfo {author} {\bibfnamefont {J.}~\bibnamefont {Grollier}},\ }\href {https://doi.org/10.1038/s42254-020-0208-2} {\bibfield  {journal} {\bibinfo  {journal} {Nature Reviews Physics}\ }\textbf {\bibinfo {volume} {2}},\ \bibinfo {pages} {499} (\bibinfo {year} {2020})}\BibitemShut {NoStop}%
\bibitem [{\citenamefont {Momeni}\ \emph {et~al.}(2025)\citenamefont {Momeni}, \citenamefont {Rahmani}, \citenamefont {Scellier}, \citenamefont {Wright}, \citenamefont {McMahon}, \citenamefont {Wanjura}, \citenamefont {Li}, \citenamefont {Skalli}, \citenamefont {Berloff}, \citenamefont {Onodera}, \citenamefont {Oguz}, \citenamefont {Morichetti}, \citenamefont {del Hougne}, \citenamefont {{Le Gallo}}, \citenamefont {Sebastian}, \citenamefont {Mirhoseini}, \citenamefont {Zhang}, \citenamefont {Markovi{\'{c}}}, \citenamefont {Brunner}, \citenamefont {Moser}, \citenamefont {Gigan}, \citenamefont {Marquardt}, \citenamefont {Ozcan}, \citenamefont {Grollier}, \citenamefont {Liu}, \citenamefont {Psaltis}, \citenamefont {Al{\`{u}}},\ and\ \citenamefont {Fleury}}]{PNNreview2025}%
  \BibitemOpen
  \bibfield  {author} {\bibinfo {author} {\bibfnamefont {A.}~\bibnamefont {Momeni}}, \bibinfo {author} {\bibfnamefont {B.}~\bibnamefont {Rahmani}}, \bibinfo {author} {\bibfnamefont {B.}~\bibnamefont {Scellier}}, \bibinfo {author} {\bibfnamefont {L.~G.}\ \bibnamefont {Wright}}, \bibinfo {author} {\bibfnamefont {P.~L.}\ \bibnamefont {McMahon}}, \bibinfo {author} {\bibfnamefont {C.~C.}\ \bibnamefont {Wanjura}}, \bibinfo {author} {\bibfnamefont {Y.}~\bibnamefont {Li}}, \bibinfo {author} {\bibfnamefont {A.}~\bibnamefont {Skalli}}, \bibinfo {author} {\bibfnamefont {N.~G.}\ \bibnamefont {Berloff}}, \bibinfo {author} {\bibfnamefont {T.}~\bibnamefont {Onodera}}, \bibinfo {author} {\bibfnamefont {I.}~\bibnamefont {Oguz}}, \bibinfo {author} {\bibfnamefont {F.}~\bibnamefont {Morichetti}}, \bibinfo {author} {\bibfnamefont {P.}~\bibnamefont {del Hougne}}, \bibinfo {author} {\bibfnamefont {M.}~\bibnamefont {{Le Gallo}}}, \bibinfo {author} {\bibfnamefont {A.}~\bibnamefont {Sebastian}}, \bibinfo {author} {\bibfnamefont
  {A.}~\bibnamefont {Mirhoseini}}, \bibinfo {author} {\bibfnamefont {C.}~\bibnamefont {Zhang}}, \bibinfo {author} {\bibfnamefont {D.}~\bibnamefont {Markovi{\'{c}}}}, \bibinfo {author} {\bibfnamefont {D.}~\bibnamefont {Brunner}}, \bibinfo {author} {\bibfnamefont {C.}~\bibnamefont {Moser}}, \bibinfo {author} {\bibfnamefont {S.}~\bibnamefont {Gigan}}, \bibinfo {author} {\bibfnamefont {F.}~\bibnamefont {Marquardt}}, \bibinfo {author} {\bibfnamefont {A.}~\bibnamefont {Ozcan}}, \bibinfo {author} {\bibfnamefont {J.}~\bibnamefont {Grollier}}, \bibinfo {author} {\bibfnamefont {A.~J.}\ \bibnamefont {Liu}}, \bibinfo {author} {\bibfnamefont {D.}~\bibnamefont {Psaltis}}, \bibinfo {author} {\bibfnamefont {A.}~\bibnamefont {Al{\`{u}}}},\ and\ \bibinfo {author} {\bibfnamefont {R.}~\bibnamefont {Fleury}},\ }\href {https://doi.org/10.1038/s41586-025-09384-2} {\bibfield  {journal} {\bibinfo  {journal} {Nature}\ }\textbf {\bibinfo {volume} {645}},\ \bibinfo {pages} {53} (\bibinfo {year} {2025})}\BibitemShut {NoStop}%
\bibitem [{\citenamefont {Palacios}\ \emph {et~al.}(2024)\citenamefont {Palacios}, \citenamefont {Mart{\'i}nez-Pe{\~n}a}, \citenamefont {Soriano}, \citenamefont {Giorgi},\ and\ \citenamefont {Zambrini}}]{Palacios2024}%
  \BibitemOpen
  \bibfield  {author} {\bibinfo {author} {\bibfnamefont {A.}~\bibnamefont {Palacios}}, \bibinfo {author} {\bibfnamefont {R.}~\bibnamefont {Mart{\'i}nez-Pe{\~n}a}}, \bibinfo {author} {\bibfnamefont {M.~C.}\ \bibnamefont {Soriano}}, \bibinfo {author} {\bibfnamefont {G.~L.}\ \bibnamefont {Giorgi}},\ and\ \bibinfo {author} {\bibfnamefont {R.}~\bibnamefont {Zambrini}},\ }\bibfield  {journal} {\bibinfo  {journal} {Communications Physics}\ }\textbf {\bibinfo {volume} {7}},\ \href {https://doi.org/10.1038/s42005-024-01859-4} {10.1038/s42005-024-01859-4} (\bibinfo {year} {2024})\BibitemShut {NoStop}%
\bibitem [{\citenamefont {Monroe}\ \emph {et~al.}(2021)\citenamefont {Monroe}, \citenamefont {Campbell}, \citenamefont {Duan}, \citenamefont {Gong}, \citenamefont {Gorshkov}, \citenamefont {Hess}, \citenamefont {Islam}, \citenamefont {Kim}, \citenamefont {Linke}, \citenamefont {Pagano}, \citenamefont {Richerme}, \citenamefont {Senko},\ and\ \citenamefont {Yao}}]{Monroe2021}%
  \BibitemOpen
  \bibfield  {author} {\bibinfo {author} {\bibfnamefont {C.}~\bibnamefont {Monroe}}, \bibinfo {author} {\bibfnamefont {W.~C.}\ \bibnamefont {Campbell}}, \bibinfo {author} {\bibfnamefont {L.-M.}\ \bibnamefont {Duan}}, \bibinfo {author} {\bibfnamefont {Z.-X.}\ \bibnamefont {Gong}}, \bibinfo {author} {\bibfnamefont {A.~V.}\ \bibnamefont {Gorshkov}}, \bibinfo {author} {\bibfnamefont {P.~W.}\ \bibnamefont {Hess}}, \bibinfo {author} {\bibfnamefont {R.}~\bibnamefont {Islam}}, \bibinfo {author} {\bibfnamefont {K.}~\bibnamefont {Kim}}, \bibinfo {author} {\bibfnamefont {N.~M.}\ \bibnamefont {Linke}}, \bibinfo {author} {\bibfnamefont {G.}~\bibnamefont {Pagano}}, \bibinfo {author} {\bibfnamefont {P.}~\bibnamefont {Richerme}}, \bibinfo {author} {\bibfnamefont {C.}~\bibnamefont {Senko}},\ and\ \bibinfo {author} {\bibfnamefont {N.~Y.}\ \bibnamefont {Yao}},\ }\href {https://doi.org/10.1103/RevModPhys.93.025001} {\bibfield  {journal} {\bibinfo  {journal} {Reviews of Modern Physics}\ }\textbf {\bibinfo {volume} {93}},\
  \bibinfo {pages} {025001} (\bibinfo {year} {2021})}\BibitemShut {NoStop}%
\bibitem [{\citenamefont {Mohseni}\ \emph {et~al.}(2022)\citenamefont {Mohseni}, \citenamefont {McMahon},\ and\ \citenamefont {Byrnes}}]{Mohseni2022}%
  \BibitemOpen
  \bibfield  {author} {\bibinfo {author} {\bibfnamefont {N.}~\bibnamefont {Mohseni}}, \bibinfo {author} {\bibfnamefont {P.~L.}\ \bibnamefont {McMahon}},\ and\ \bibinfo {author} {\bibfnamefont {T.}~\bibnamefont {Byrnes}},\ }\href {https://doi.org/10.1038/s42254-022-00440-8} {\bibfield  {journal} {\bibinfo  {journal} {Nature Reviews Physics}\ }\textbf {\bibinfo {volume} {4}},\ \bibinfo {pages} {363} (\bibinfo {year} {2022})}\BibitemShut {NoStop}%
\bibitem [{\citenamefont {Vandoorne}\ \emph {et~al.}(2014)\citenamefont {Vandoorne}, \citenamefont {Mechet}, \citenamefont {Van~Vaerenbergh}, \citenamefont {Fiers}, \citenamefont {Morthier}, \citenamefont {Verstraeten}, \citenamefont {Schrauwen}, \citenamefont {Dambre},\ and\ \citenamefont {Bienstman}}]{Vandoorne2014}%
  \BibitemOpen
  \bibfield  {author} {\bibinfo {author} {\bibfnamefont {K.}~\bibnamefont {Vandoorne}}, \bibinfo {author} {\bibfnamefont {P.}~\bibnamefont {Mechet}}, \bibinfo {author} {\bibfnamefont {T.}~\bibnamefont {Van~Vaerenbergh}}, \bibinfo {author} {\bibfnamefont {M.}~\bibnamefont {Fiers}}, \bibinfo {author} {\bibfnamefont {G.}~\bibnamefont {Morthier}}, \bibinfo {author} {\bibfnamefont {D.}~\bibnamefont {Verstraeten}}, \bibinfo {author} {\bibfnamefont {B.}~\bibnamefont {Schrauwen}}, \bibinfo {author} {\bibfnamefont {J.}~\bibnamefont {Dambre}},\ and\ \bibinfo {author} {\bibfnamefont {P.}~\bibnamefont {Bienstman}},\ }\href {https://doi.org/10.1038/ncomms4541} {\bibfield  {journal} {\bibinfo  {journal} {Nature Communications}\ }\textbf {\bibinfo {volume} {5}},\ \bibinfo {pages} {3541} (\bibinfo {year} {2014})}\BibitemShut {NoStop}%
\bibitem [{\citenamefont {Donelli}\ \emph {et~al.}(2025)\citenamefont {Donelli}, \citenamefont {Gherardini}, \citenamefont {Marino}, \citenamefont {Campaioli},\ and\ \citenamefont {Buffoni}}]{Donelli2025}%
  \BibitemOpen
  \bibfield  {author} {\bibinfo {author} {\bibfnamefont {B.}~\bibnamefont {Donelli}}, \bibinfo {author} {\bibfnamefont {S.}~\bibnamefont {Gherardini}}, \bibinfo {author} {\bibfnamefont {R.}~\bibnamefont {Marino}}, \bibinfo {author} {\bibfnamefont {F.}~\bibnamefont {Campaioli}},\ and\ \bibinfo {author} {\bibfnamefont {L.}~\bibnamefont {Buffoni}},\ }\href {https://doi.org/10.1103/lffq-ylgz} {\bibfield  {journal} {\bibinfo  {journal} {Phys. Rev. E}\ }\textbf {\bibinfo {volume} {111}},\ \bibinfo {pages} {L062102} (\bibinfo {year} {2025})}\BibitemShut {NoStop}%
\bibitem [{\citenamefont {Montangero}(2018)}]{Montangero2018}%
  \BibitemOpen
  \bibfield  {author} {\bibinfo {author} {\bibfnamefont {S.}~\bibnamefont {Montangero}},\ }\href {https://doi.org/10.1007/978-3-030-01409-4} {\emph {\bibinfo {title} {Introduction to Tensor Network Methods}}},\ \bibinfo {edition} {1st}\ ed.\ (\bibinfo  {publisher} {Springer Cham},\ \bibinfo {year} {2018})\BibitemShut {NoStop}%
\bibitem [{\citenamefont {Gravina}\ \emph {et~al.}(2025)\citenamefont {Gravina}, \citenamefont {Savona},\ and\ \citenamefont {Vicentini}}]{Gravina2025}%
  \BibitemOpen
  \bibfield  {author} {\bibinfo {author} {\bibfnamefont {L.}~\bibnamefont {Gravina}}, \bibinfo {author} {\bibfnamefont {V.}~\bibnamefont {Savona}},\ and\ \bibinfo {author} {\bibfnamefont {F.}~\bibnamefont {Vicentini}},\ }\href {https://doi.org/10.22331/q-2025-07-22-1803} {\bibfield  {journal} {\bibinfo  {journal} {Quantum}\ }\textbf {\bibinfo {volume} {9}},\ \bibinfo {pages} {1803} (\bibinfo {year} {2025})}\BibitemShut {NoStop}%
\bibitem [{\citenamefont {Sinibaldi}\ \emph {et~al.}(2026)\citenamefont {Sinibaldi}, \citenamefont {Hendry}, \citenamefont {Vicentini},\ and\ \citenamefont {Carleo}}]{Sinibaldi2026}%
  \BibitemOpen
  \bibfield  {author} {\bibinfo {author} {\bibfnamefont {A.}~\bibnamefont {Sinibaldi}}, \bibinfo {author} {\bibfnamefont {D.}~\bibnamefont {Hendry}}, \bibinfo {author} {\bibfnamefont {F.}~\bibnamefont {Vicentini}},\ and\ \bibinfo {author} {\bibfnamefont {G.}~\bibnamefont {Carleo}},\ }\href {https://doi.org/10.1103/kqvx-dl54} {\bibfield  {journal} {\bibinfo  {journal} {Phys. Rev. Lett.}\ }\textbf {\bibinfo {volume} {136}},\ \bibinfo {pages} {120402} (\bibinfo {year} {2026})}\BibitemShut {NoStop}%
\bibitem [{\citenamefont {Luchnikov}\ \emph {et~al.}(2019)\citenamefont {Luchnikov}, \citenamefont {Vintskevich}, \citenamefont {Ouerdane},\ and\ \citenamefont {Filippov}}]{Luchnikov2019}%
  \BibitemOpen
  \bibfield  {author} {\bibinfo {author} {\bibfnamefont {I.~A.}\ \bibnamefont {Luchnikov}}, \bibinfo {author} {\bibfnamefont {S.~V.}\ \bibnamefont {Vintskevich}}, \bibinfo {author} {\bibfnamefont {H.}~\bibnamefont {Ouerdane}},\ and\ \bibinfo {author} {\bibfnamefont {S.~N.}\ \bibnamefont {Filippov}},\ }\href {https://doi.org/10.1103/PhysRevLett.122.160401} {\bibfield  {journal} {\bibinfo  {journal} {Physical Review Letters}\ }\textbf {\bibinfo {volume} {122}},\ \bibinfo {pages} {160401} (\bibinfo {year} {2019})}\BibitemShut {NoStop}%
\bibitem [{\citenamefont {Ghosh}\ \emph {et~al.}(2019)\citenamefont {Ghosh}, \citenamefont {Opala}, \citenamefont {Matuszewski}, \citenamefont {Paterek},\ and\ \citenamefont {Liew}}]{Ghosh2019}%
  \BibitemOpen
  \bibfield  {author} {\bibinfo {author} {\bibfnamefont {S.}~\bibnamefont {Ghosh}}, \bibinfo {author} {\bibfnamefont {A.}~\bibnamefont {Opala}}, \bibinfo {author} {\bibfnamefont {M.}~\bibnamefont {Matuszewski}}, \bibinfo {author} {\bibfnamefont {T.}~\bibnamefont {Paterek}},\ and\ \bibinfo {author} {\bibfnamefont {T.~C.~H.}\ \bibnamefont {Liew}},\ }\href {https://doi.org/10.1038/s41534-019-0149-8} {\bibfield  {journal} {\bibinfo  {journal} {npj Quantum Information}\ }\textbf {\bibinfo {volume} {5}},\ \bibinfo {pages} {35} (\bibinfo {year} {2019})}\BibitemShut {NoStop}%
\bibitem [{\citenamefont {Tran}\ and\ \citenamefont {Nakajima}(2021)}]{TranNakajima2021}%
  \BibitemOpen
  \bibfield  {author} {\bibinfo {author} {\bibfnamefont {Q.~H.}\ \bibnamefont {Tran}}\ and\ \bibinfo {author} {\bibfnamefont {K.}~\bibnamefont {Nakajima}},\ }\href {https://doi.org/10.1103/PhysRevLett.127.260401} {\bibfield  {journal} {\bibinfo  {journal} {Physical Review Letters}\ }\textbf {\bibinfo {volume} {127}},\ \bibinfo {pages} {260401} (\bibinfo {year} {2021})}\BibitemShut {NoStop}%
\bibitem [{\citenamefont {Hoerl}\ and\ \citenamefont {Kennard}(1970)}]{Hoerl1970}%
  \BibitemOpen
  \bibfield  {author} {\bibinfo {author} {\bibfnamefont {A.~E.}\ \bibnamefont {Hoerl}}\ and\ \bibinfo {author} {\bibfnamefont {R.~W.}\ \bibnamefont {Kennard}},\ }\href {https://doi.org/10.1080/00401706.1970.10488634} {\bibfield  {journal} {\bibinfo  {journal} {Technometrics}\ }\textbf {\bibinfo {volume} {12}},\ \bibinfo {pages} {55} (\bibinfo {year} {1970})}\BibitemShut {NoStop}%
\bibitem [{\citenamefont {Trouvain}\ \emph {et~al.}(2020)\citenamefont {Trouvain}, \citenamefont {Pedrelli}, \citenamefont {Dinh},\ and\ \citenamefont {Hinaut}}]{Trouvain2020}%
  \BibitemOpen
  \bibfield  {author} {\bibinfo {author} {\bibfnamefont {N.}~\bibnamefont {Trouvain}}, \bibinfo {author} {\bibfnamefont {L.}~\bibnamefont {Pedrelli}}, \bibinfo {author} {\bibfnamefont {T.~T.}\ \bibnamefont {Dinh}},\ and\ \bibinfo {author} {\bibfnamefont {X.}~\bibnamefont {Hinaut}},\ }in\ \href {https://doi.org/10.1007/978-3-030-61616-8_40} {\emph {\bibinfo {booktitle} {Artificial Neural Networks and Machine Learning {\textendash} {ICANN} 2020}}}\ (\bibinfo  {publisher} {Springer International Publishing},\ \bibinfo {year} {2020})\ pp.\ \bibinfo {pages} {494--505}\BibitemShut {NoStop}%
\bibitem [{\citenamefont {Hoeffding}(1963)}]{hoeffding1963probability}%
  \BibitemOpen
  \bibfield  {author} {\bibinfo {author} {\bibfnamefont {W.}~\bibnamefont {Hoeffding}},\ }\href {https://doi.org/10.1080/01621459.1963.10500830} {\bibfield  {journal} {\bibinfo  {journal} {Journal of the American Statistical Association}\ }\textbf {\bibinfo {volume} {58}},\ \bibinfo {pages} {13} (\bibinfo {year} {1963})}\BibitemShut {NoStop}%
\bibitem [{\citenamefont {Holzer}\ and\ \citenamefont {Turkalj}(2026)}]{holzer2026spectral}%
  \BibitemOpen
  \bibfield  {author} {\bibinfo {author} {\bibfnamefont {P.}~\bibnamefont {Holzer}}\ and\ \bibinfo {author} {\bibfnamefont {I.}~\bibnamefont {Turkalj}},\ }\href {https://doi.org/10.1007/s42484-026-00372-x} {\bibfield  {journal} {\bibinfo  {journal} {Quantum Machine Intelligence}\ }\textbf {\bibinfo {volume} {8}},\ \bibinfo {pages} {33} (\bibinfo {year} {2026})}\BibitemShut {NoStop}%
\bibitem [{\citenamefont {Sidon}(1932)}]{sidon1932satz}%
  \BibitemOpen
  \bibfield  {author} {\bibinfo {author} {\bibfnamefont {S.}~\bibnamefont {Sidon}},\ }\href {https://doi.org/10.1007/BF01455900} {\bibfield  {journal} {\bibinfo  {journal} {Mathematische Annalen}\ }\textbf {\bibinfo {volume} {106}},\ \bibinfo {pages} {536} (\bibinfo {year} {1932})}\BibitemShut {NoStop}%
\bibitem [{\citenamefont {Mart{\'i}nez-Pe{\~n}a}\ \emph {et~al.}(2023)\citenamefont {Mart{\'i}nez-Pe{\~n}a}, \citenamefont {Nokkala}, \citenamefont {Giorgi}, \citenamefont {Zambrini},\ and\ \citenamefont {Soriano}}]{MartinezPena2023IPC}%
  \BibitemOpen
  \bibfield  {author} {\bibinfo {author} {\bibfnamefont {R.}~\bibnamefont {Mart{\'i}nez-Pe{\~n}a}}, \bibinfo {author} {\bibfnamefont {J.}~\bibnamefont {Nokkala}}, \bibinfo {author} {\bibfnamefont {G.~L.}\ \bibnamefont {Giorgi}}, \bibinfo {author} {\bibfnamefont {R.}~\bibnamefont {Zambrini}},\ and\ \bibinfo {author} {\bibfnamefont {M.~C.}\ \bibnamefont {Soriano}},\ }\href {https://doi.org/10.1007/s12559-020-09772-y} {\bibfield  {journal} {\bibinfo  {journal} {Cognitive Computation}\ }\textbf {\bibinfo {volume} {15}},\ \bibinfo {pages} {1440} (\bibinfo {year} {2023})}\BibitemShut {NoStop}%
\bibitem [{\citenamefont {Jaeger}(2001)}]{jaeger2001echo}%
  \BibitemOpen
  \bibfield  {author} {\bibinfo {author} {\bibfnamefont {H.}~\bibnamefont {Jaeger}},\ }\href {https://www.ai.rug.nl/minds/uploads/EchoStatesTechRep.pdf} {\emph {\bibinfo {title} {The ``Echo State'' Approach to Analysing and Training Recurrent Neural Networks -- with an Erratum Note}}},\ \bibinfo {type} {Tech. Rep.}\ \bibinfo {number} {148}\ (\bibinfo  {institution} {German National Research Center for Information Technology GMD},\ \bibinfo {address} {Bonn, Germany},\ \bibinfo {year} {2001})\BibitemShut {NoStop}%
\bibitem [{\citenamefont {Lindblad}(1976)}]{lindblad1976}%
  \BibitemOpen
  \bibfield  {author} {\bibinfo {author} {\bibfnamefont {G.}~\bibnamefont {Lindblad}},\ }\href {https://doi.org/10.1007/BF01608499} {\bibfield  {journal} {\bibinfo  {journal} {Communications in Mathematical Physics}\ }\textbf {\bibinfo {volume} {48}},\ \bibinfo {pages} {119} (\bibinfo {year} {1976})}\BibitemShut {NoStop}%
\bibitem [{\citenamefont {Bayraktar}\ \emph {et~al.}(2023)\citenamefont {Bayraktar}, \citenamefont {Charara}, \citenamefont {Clark}, \citenamefont {Cohen}, \citenamefont {Costa}, \citenamefont {Fang}, \citenamefont {Gao}, \citenamefont {Guan}, \citenamefont {Gunnels}, \citenamefont {Haidar} \emph {et~al.}}]{bayraktar2023cuquantum}%
  \BibitemOpen
  \bibfield  {author} {\bibinfo {author} {\bibfnamefont {H.}~\bibnamefont {Bayraktar}}, \bibinfo {author} {\bibfnamefont {A.}~\bibnamefont {Charara}}, \bibinfo {author} {\bibfnamefont {D.}~\bibnamefont {Clark}}, \bibinfo {author} {\bibfnamefont {S.}~\bibnamefont {Cohen}}, \bibinfo {author} {\bibfnamefont {T.}~\bibnamefont {Costa}}, \bibinfo {author} {\bibfnamefont {Y.-L.~L.}\ \bibnamefont {Fang}}, \bibinfo {author} {\bibfnamefont {Y.}~\bibnamefont {Gao}}, \bibinfo {author} {\bibfnamefont {J.}~\bibnamefont {Guan}}, \bibinfo {author} {\bibfnamefont {J.}~\bibnamefont {Gunnels}}, \bibinfo {author} {\bibfnamefont {A.}~\bibnamefont {Haidar}}, \emph {et~al.},\ }in\ \href {https://doi.org/10.1109/QCE57702.2023.00119} {\emph {\bibinfo {booktitle} {2023 IEEE International Conference on Quantum Computing and Engineering (QCE)}}},\ Vol.~\bibinfo {volume} {1}\ (\bibinfo {organization} {IEEE},\ \bibinfo {year} {2023})\ pp.\ \bibinfo {pages} {1050--1061}\BibitemShut {NoStop}%
\bibitem [{\citenamefont {Suzuki}(1976)}]{suzuki1976relationship}%
  \BibitemOpen
  \bibfield  {author} {\bibinfo {author} {\bibfnamefont {M.}~\bibnamefont {Suzuki}},\ }\href {https://doi.org/10.1143/PTP.56.1454} {\bibfield  {journal} {\bibinfo  {journal} {Progress of Theoretical Physics}\ }\textbf {\bibinfo {volume} {56}},\ \bibinfo {pages} {1454} (\bibinfo {year} {1976})}\BibitemShut {NoStop}%
\bibitem [{\citenamefont {{Pawsey Supercomputing Research Centre}}(2026)}]{pawsey_setonixq_quickstart}%
  \BibitemOpen
  \bibfield  {author} {\bibinfo {author} {\bibnamefont {{Pawsey Supercomputing Research Centre}}},\ }\href {https://pawsey.atlassian.net/wiki/spaces/US/pages/1063682139} {\bibinfo {title} {Quantum partition: Quick start}},\ \bibinfo {howpublished} {Pawsey User Support Documentation} (\bibinfo {year} {2026}),\ \bibinfo {note} {accessed 15 August 2026}\BibitemShut {NoStop}%
\end{thebibliography}%

\end{document}